\documentclass[12pt]{article}
\pdfoutput=1
\usepackage{hyperref}
\usepackage{graphicx}
\usepackage{graphics}
\usepackage{dsfont}
\usepackage{epsfig}
\usepackage{amsmath,amssymb,amsthm,amscd}
\usepackage{simpler-wick}
\usepackage {subcaption}

\def\ket{\rangle}
\def\bra{\langle}

\newcommand{\be}{\begin{equation}}
\newcommand{\ee}{\end{equation}}
\newcommand{\ba}{\begin{aligned}}
\newcommand{\ea}{\end{aligned}}

\numberwithin{equation}{section}

\begin{document}
\begin{titlepage}

\rightline{USTC-ICTS/PCFT-21-43}

\vskip 3 cm

\centerline{\Large 
\bf  
Bootstrapping Calabi-Yau  Quantum Mechanics  }

\vskip 0.5 cm

\renewcommand{\thefootnote}{\fnsymbol{footnote}}
\vskip 30pt \centerline{ {\large \rm 
Bao-ning Du\footnote{baoningd@mail.ustc.edu.cn},
Min-xin Huang\footnote{minxin@ustc.edu.cn},   Pei-xuan Zeng\footnote{zengpx@mail.ustc.edu.cn}
} } \vskip .5cm  \vskip 20pt 

\begin{center}
{Interdisciplinary Center for Theoretical Study,  \\ \vskip 0.1cm  University of Science and Technology of China,  Hefei, Anhui 230026, China} 
 \\ \vskip 0.3 cm
{Peng Huanwu Center for Fundamental Theory,  \\ \vskip 0.1cm  Hefei, Anhui 230026, China} 
\end{center}

\setcounter{footnote}{0}
\renewcommand{\thefootnote}{\arabic{footnote}}
\vskip 40pt
\begin{abstract}

Recently, a novel bootstrap method for numerical calculations in matrix models and quantum mechanical systems is proposed.   We apply the method to certain quantum mechanical systems derived from some well-known local toric Calabi-Yau geometries, where the exact quantization conditions have been conjecturally related to topological string theory. We find that the bootstrap method provides a promising alternative for the precision numerical calculations of the energy eigenvalues. An improvement in our approach is to use a larger set of two-dimensional operators instead of one-dimensional ones. We also apply our improved bootstrap methods to some non-relativistic models in the recent literature and demonstrate better numerical accuracies.

\end{abstract}

\end{titlepage}
\vfill \eject


\newpage

\baselineskip=16pt

\tableofcontents

\section{Introduction}

Recently, a novel bootstrap method for numerical calculations in matrix models \cite{Lin:2020mme} and matrix quantum mechanics \cite{Han:2020bkb} is proposed, and has been studied in the literature \cite{Kazakov:2021lel, Berenstein:2021dyf, Berenstein:2021loy, Bhattacharya:2021btd, Aikawa:2021eai, Aikawa:2021qbl}. This is related to and inspired by early works on matrix models, see e.g. the recent paper \cite{Koch:2021yeb, Jha:2021exo}. In particular, the method appears  promising for precise numerical calculations of energy eigenvalues in quantum mechanical systems, which are essential for testing exact quantization conditions. 

The studies of exact quantization conditions including non-perturbative contributions, e.g. of the non-analytic form $e^{-\frac{A}{\hbar}}$ from instantons, have a long history, see e.g. early works \cite{Bender:1969si, Zinn-Justin:1981qzi}.  The resurgent methods provide a framework for mathematically rigorous proofs of such exact quantization conditions, see e.g. \cite{AIHPA_1999__71_1_1_0}. The conventional Hamiltonians of a one-dimensional non-relativistic particle with general polynomial potentials have been much well studied, and the exact quantization conditions are most recently derived in terms of  TBA (Thermodynamic Bethe Ansatz) equations \cite{Ito:2018eon, Gabai:2021dfi}. One can consider more general quantum mechanical systems. Nekrasov and Shatashvili proposed the exact quantization conditions for certain integrable systems using the  Nekrasov partition function of four dimensional $\mathcal{N}=2$ supersymmetric Seiberg-Witten gauge theories \cite{Nekrasov:2002qd, Nekrasov:2009rc}. In a closely related setting, Grassi and Mari\~no considered a class of Hamiltonians with a deformed kinetic term and polynomial potentials, i.e. $\hat{H} = \cosh(\hat{p}) + P(\hat{x})$ \cite{Grassi:2018bci}.

In this paper, we apply the bootstrap method to the class of quantum mechanics systems derived from the mirror curves of some well known local toric Calabi-Yau geometries, where the Hamiltonians are exponential functions of both canonical position and momentum operators. The relations between quantum periods and TBA-like equations for these Calabi-Yau geometries are also recently studied in \cite{Du:2020nwl}. The quantization of mirror curves and the relation to topological string theory have been long considered e.g. in the pioneering papers \cite{Aganagic:2003qj, Aganagic:2011mi}. Inspired by the precision numerical calculations of the spectra \cite{Huang:2014eha}, the exact quantization conditions are conjectured in \cite{Grassi:2014zfa, Wang:2015wdy} using refined topological string theory, often related to the partition functions of five dimensional supersymmetric gauge theories. Despite many tests, to our knowledge,  except for some cases with special  values of the Planck constant  in e.g. \cite{Kashaev:2017zmv, Kashaev:2019gkn}, the proposals in \cite{Grassi:2014zfa, Wang:2015wdy} and the subsequent generalizations to higher genus mirror curves remain largely conjectural. Thus it is helpful to develop novel tools for numerical calculations, for the purposes of potentially more precise tests of the exact quantizations as well as obtaining new results in other less explored quantum mechanical systems.    

In the Calabi-Yau models, it is natural to treat the momentum and position operators equally, and the bootstrap operator is a linear combination of operators in a two-dimensional space indexed by both momentum and position, instead of the one-dimensional space in the literature  \cite{Han:2020bkb, Berenstein:2021dyf, Berenstein:2021loy, Bhattacharya:2021btd, Aikawa:2021eai, Aikawa:2021qbl}. This turns out to also work for the more conventional quantum mechanical models with the standard non-relativistic kinetic term, and improves the bootstrap efficiency. For completeness and comparisons with the Calabi-Yau models, we also consider some of these non-relativistic models to demonstrate the improvements.  

The paper is organized as the followings. We study two simple Calabi-Yau models in Sec. \ref{P1P1} and Sec. \ref{P2}, and compare the numerical precision of bootstrap method and the previous conventional method in \cite{Huang:2014eha}, which uses a basis of the energy eigenfunctions of a harmonic oscillator and truncate to a finite level for the numerical calculations. We will refer to this competing method as the ``truncation method". We give more details of the formalism as applied to our case  for the first  example of the $\mathbb{P}^1\times \mathbb{P}^1$ model, which has a large symmetry and also belongs to the class of integrable systems known as the relativistic Toda models. The calculations would be rather similar for the other $\mathbb{P}^2$ model. In Sec. \ref{SU2} we study a non-relativistic Toda model with exponential potential, which is related to the four dimensional pure $SU(2)$  Seiberg-Witten theory  \cite{Nekrasov:2009rc, Grassi:2018bci}. In Sec. \ref{secquartic} we revisit the well known quartic harmonic oscillator model, apply our improved bootstrap method with the two-dimensional operators, and demonstrate better numerical accuracies than those in the literature. We give a summary and some potential future directions in the conclusion in Sec. \ref{conclusion}.


\section{The $\mathbb{P}^1\times \mathbb{P}^1$ model} 
\label{P1P1}

The Hamiltonian is 
\be \label{P1P1H}
\hat{H} = e^{\hat{x}} + e^{- \hat{x}} + e^{\hat{p}} + e^{-\hat{p}}. 
\ee
We note that the notation is a bit different from  e.g. \cite{Huang:2014eha}, which identifies the above expression as the exponential of the Hamiltonian as it is more convenient in the context of topological string theory.  In the current notation, the inverse Hamiltonian $\hat{H}^{-1}$ is a trace class operator and has a mathematically well-defined discrete spectrum. We consider the expectation values of the operators $e^{m \hat{x} +n \hat{p}}$ in an energy eigenstate, and denote 
\be 
f_{m,n} := \bra e^{m \hat{x} +n \hat{p}} \ket  .
\ee
We use a properly normalized state so $f_{0,0}=1$. Since the canonical commutation relation $[\hat{x}, \hat{p}] =i\hbar $ and the Hamiltonian are unchanged under the symplectic transformations $(x,p)\rightarrow (-p,x), (p, -x) $, we have the apparent symmetry property 
\be
f_{m,n} = f_{-n,m} = f_{n, -m} =f_{-m,-n} .  
\ee
Furthermore, we also have an additional symmetry $f_{m,n} = f_{n,m}$, which amounts to an exchange of $x,p$ or switching the sign of the Planck constant $\hbar\rightarrow -\hbar$. To see this, we note that the Hamiltonian is invariant under the $\mathcal{T}$ symmetry which replaces $p\rightarrow -p, x\rightarrow x, i\rightarrow -i$ or maps a wave function in position space to its complex conjugate, so we can always choose the energy eigenfunctions to be real functions of $x$. The combined $\mathcal{P} \mathcal{T}$ symmetry is well studied in the context of non-Hermitian Hamiltonians with real spectra \cite{Bender:1998ke}. We can write  the expectation value in terms of the real wave function 
\be 
f_{m,n} = \int_{-\infty} ^{\infty} dx~ \psi(x) e^{mx- \frac{mn}{2} i\hbar }  \psi(x-in \hbar).
\ee
The integral is apparently invariant under $\hbar\rightarrow -\hbar$ by a shift of the integration variable $x\rightarrow x+in\hbar$. Our assumption here is that the wave function $\psi(x)$ can be analytically continued to the $x$ complex plane, and there is no singularity inside the rectangle with edges $[-R, R], [R,R+in\hbar], [R+in\hbar, -R+in\hbar],  [-R+in\hbar, -R]$ for any $R>0$ so that its contour integral vanishes. Furthermore, the wave function is normalizable (square integrable) so the integrals along the two edges  $[R,R+in\hbar],   [-R+in\hbar, -R]$ vanish as $R\rightarrow +\infty$. So overall we have a very large symmetry which shall much simplify the calculations  
 \be \label{symmetry} 
f_{m,n} = f_{n,m} = f_{|m|, |n|}.  
\ee
In the followings we will assume $\hbar$ is a real positive constant. 

We apply the bootstrap relations 
\be \label{bootstrap relations}
\bra \hat{H}  e^{m \hat{x} +n \hat{p}} \ket  = \bra e^{m \hat{x} +n \hat{p}} \hat{H} \ket = E  \bra e^{m \hat{x} +n \hat{p}} \ket ,
\ee
where $E$ is the energy eigenvalue. After applying the well-known Baker-Campbell-Hausdorff formula, this gives some recursion relations among the expectation values of operators 
\be \ba  \label{recursion}
\sin(\frac{n\hbar}{2}) (f_{m+1,n} -f_{m-1,n}) + \sin(\frac{ m \hbar}{2}) (f_{m,n-1} -f_{m,n+1}) &=  0, \\
\cos(\frac{n\hbar}{2}) (f_{m+1,n} +f_{m-1,n}) + \cos(\frac{ m \hbar}{2}) (f_{m,n-1} + f_{m,n+1})   &= Ef_{m,n}.
\ea  \ee 
For example, with the symmetry (\ref{symmetry}) it is easy to see $E=4f_{1,0}$. These recursion relations are not completely independent, as one can check that some are simply related to the symmetry relations (\ref{symmetry}). It turns out that besides the energy $E$, we need one more initial condition to start the bootstrap to compute the $f_{m,n}$'s for all integers $m,n$. A convenient choice is to use $f_{2,1}$ as the initial condition. 

Consider the operator 
\be \label{operator2.8}
\mathbb{O} = \sum_{m,n} c_{m,n} e^{m \hat{x} +n \hat{p}}. 
\ee 
The positivity of the expectation value $\bra \mathbb{O} ^{\dagger} \mathbb{O} \ket $ with any coefficients  is equivalent to the positivity of the Hermitian matrix 
\be \label{matrix2.9}
M_{(m,n), (m^\prime, n^{\prime})} =e^{\frac{i\hbar}{2} (m n^{\prime} -n m^{\prime}) }   f_{m+m^\prime, n+n^\prime}  . 
\ee
We can choose the sum in the operator $\mathbb{O}$ over a finite set, and use the positivity of the above matrix to constrain the energy eigenvalue $E$ and the initial value $f_{2,1}$. We refer to the matrix (\ref{matrix2.9}) as the ``bootstrap matrix" and note that in our case it is not necessarily real and symmetric as in the cases of simple non-relativistic quantum systems in e.g. \cite{Berenstein:2021dyf, Berenstein:2021loy}. For such a matrix, we refer to the number $K\equiv \max\{|m|, |n|, |m^{\prime}|, |n^{\prime}| \}$ as the bootstrap level.

There is a caveat with bootstrap method for our model that we can not arbitrarily increase the bootstrap level $K$ in the operator (\ref{operator2.8}). In a harmonic oscillator with unit mass and frequency, it is well known that the normalizable wave functions have the asymptotic behavior $\psi(x) \sim e^{-\frac{x^2}{2\hbar} }$ as $x\sim \pm \infty$, so the expectation value of $e^{m \hat{x} +n \hat{p}}$ is always finite. However, the asymptotic behavior of the energy eigenfunctions in our model is different. For small $\hbar$, it is easy to solve the leading order WKB wave function. Denote the ansatz for the wave function 
\be 
\psi(x) = \exp[\int^x \frac{i}{\hbar} w(x^\prime ) dx^\prime ]. 
\ee
The leading order equation for $w(x)$ is obtained by simply replacing $\hat{p}$ with $w(x)$ in the Hamiltonian 
\be 
 e^{x} + e^{- x} + e^{w(x)} + e^{- w(x) } = E . 
 \ee
The asymptotic behavior of the wave function is determined by the imaginary part of $w(x)$ as $x\sim \pm \infty$. It is easy the solve the equation and find 
\be 
\Im (w(x) )= \log(-1) +\mathcal{O} (e^{-|x|}). 
\ee 
As similar to the case of harmonic oscillator, the asymptotic behavior of the normalizable wave function in our model for small $\hbar$ comes from the slowest decaying branch, i.e. we have  
\be \label{asymp2.13}
\psi(x) \sim e^{- \frac{\pi}{\hbar} |x| }, ~~~~  x\sim \pm \infty. 
\ee
This constrains the operators that we can put in the sum in   (\ref{operator2.8}). For example, the expectation value of $e^{mx}$ for $|m|> \frac{2\pi}{\hbar}$ is infinite as the integral diverges,  thus is not admissible for bootstrap. More detailed properties of the wave function in this model were studied in e.g. \cite{Marino:2016rsq, Zakany:2017txl}.

We review some details of the direct computations in \cite{Huang:2014eha} with the ``truncation method", which use the basis of the wave eigenfunctions of the quantum harmonic oscillator with unit mass and frequency 
\begin{eqnarray} \label{wave2.14}
\psi_n(x) =\frac{1}{\sqrt{2^n n!}} \left( \frac{1}{\pi \hbar}\right) ^{\frac{1}{4}} e^{-\frac{ x^2}{2\hbar}} H_n (\frac{x}{\sqrt{\hbar}} ),~~~n\geq 0,
\end{eqnarray}
where $H_n(x)$ are the Hermite polynomials. There is a useful integral 
\begin{eqnarray} 
\int_{-\infty}^{\infty} e^{-x^2} H_{m}(x+y) H_{n}(x+z) dx =2^{n} \sqrt{\pi} m! z^{n-m} L^{n-m}_{m} (-2yz), ~~ m \leq n,
\end{eqnarray}
where $L^\alpha_n(z)$ are the Laguerre polynomials. Up to a finite level, we can numerically compute the matrix elements of the Hamiltonian $\hat{H}$ in the harmonic oscillator basis, and diagonalize to find the energy eigenvalues and eigenstates. We can then compute the expectation value $f_{m,n}$ for a particular energy eigenstate. For example, we have the matrix elements 
\be
\bra \psi_{m } | e^{k x} |\psi_{n} \ket = \sqrt{\frac{m!}{n!}} (\frac{ k^2 \hbar}{2})^{\frac{n-m}{2}} e^{\frac{k^2\hbar}{4}}  
L^{n-m}_{m}  (-\frac{k^2\hbar}{2} ) ,~~ m \leq n,
\ee 
where the cases of $m>n$ are related by complex conjugation, which does not change real matrix elements as in this case. 

\begin{figure}
\centering
	\begin{subfigure}{0.49\textwidth}
		\includegraphics[width=\textwidth]{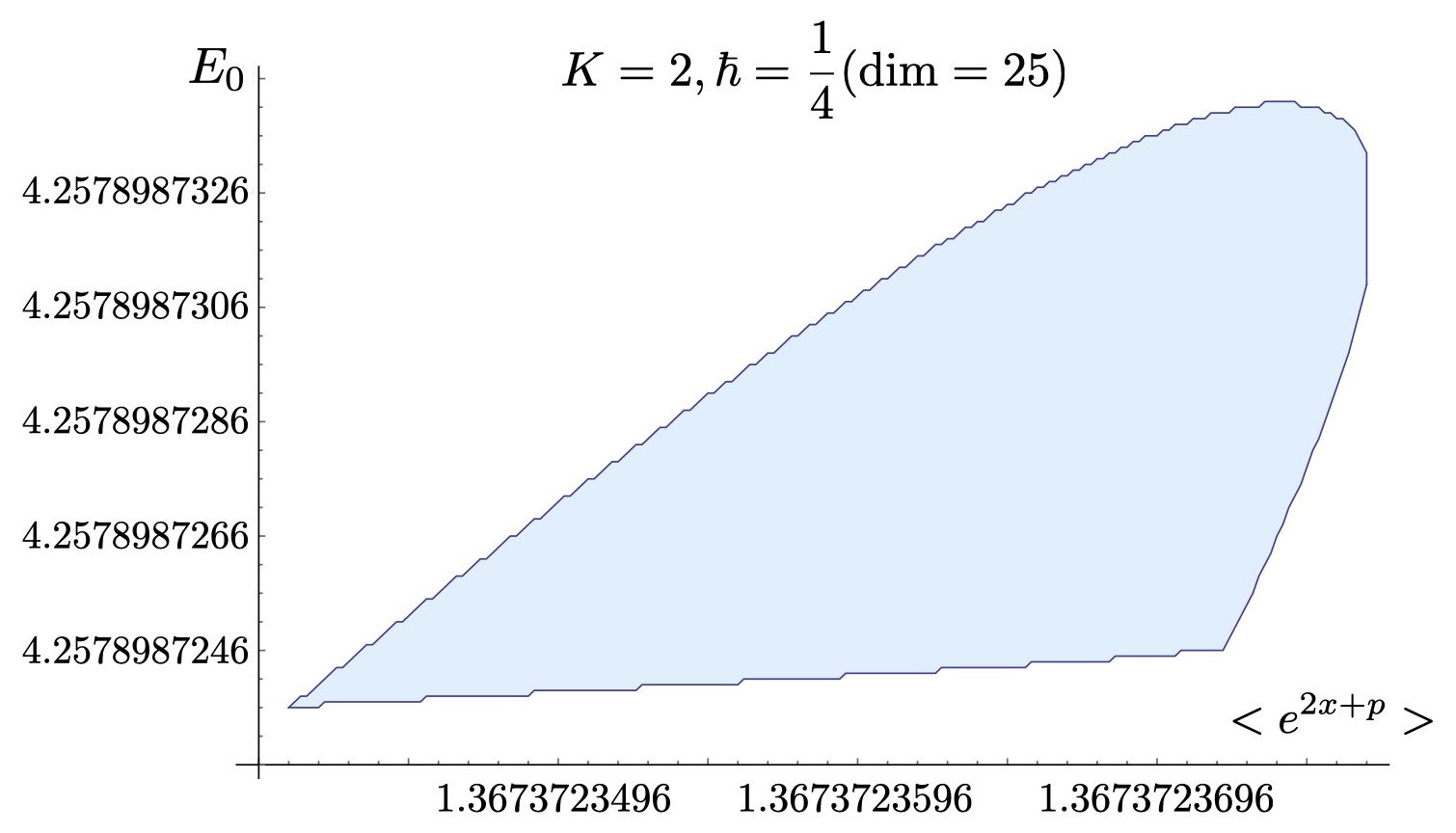} 	
	\end{subfigure}
	\begin{subfigure}{0.49\textwidth}
		\includegraphics[width=\textwidth]{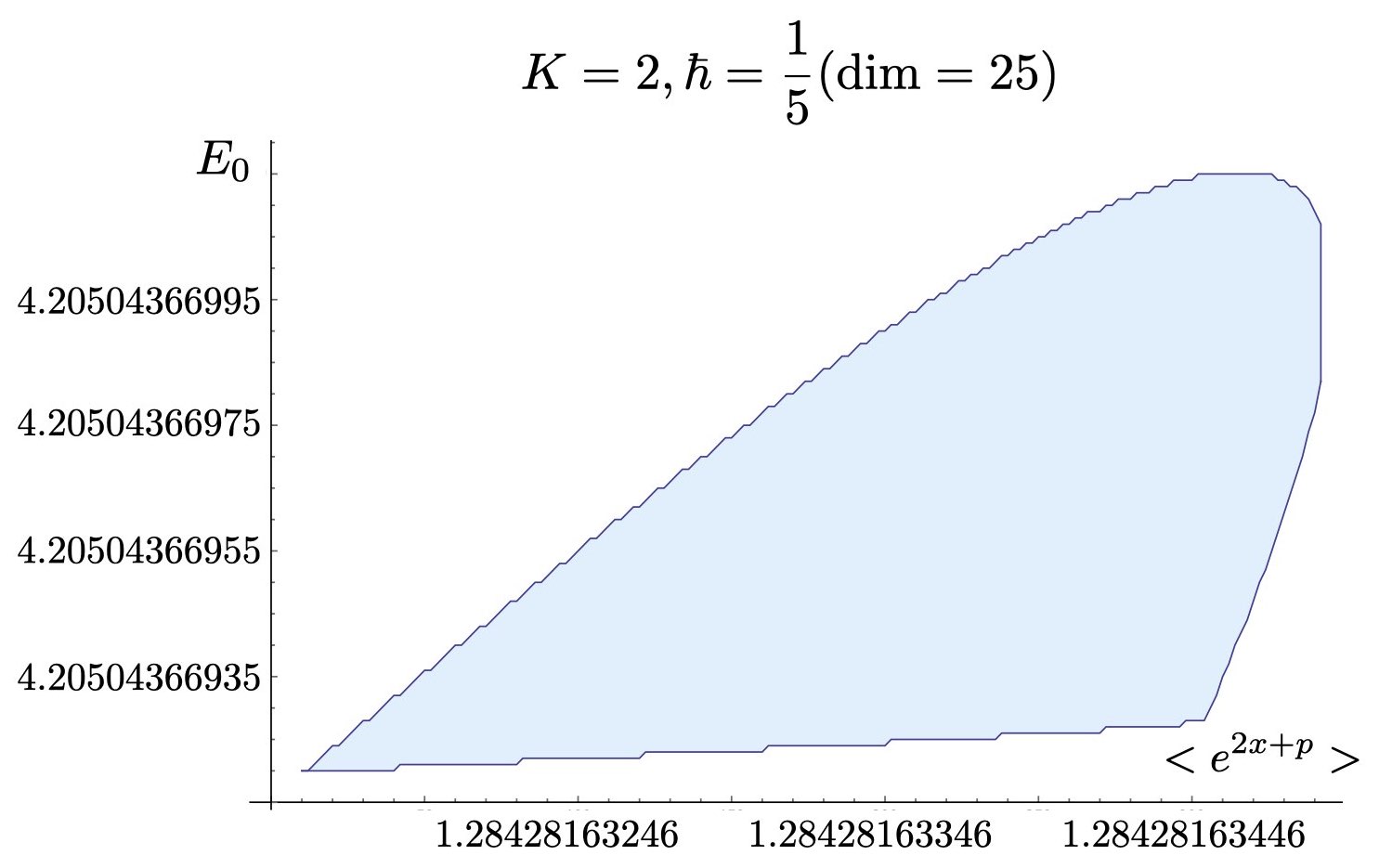}		
	\end{subfigure}
	
	 \vskip 18pt
	 
	\begin{subfigure}{0.49\textwidth}
		\includegraphics[width=\textwidth]{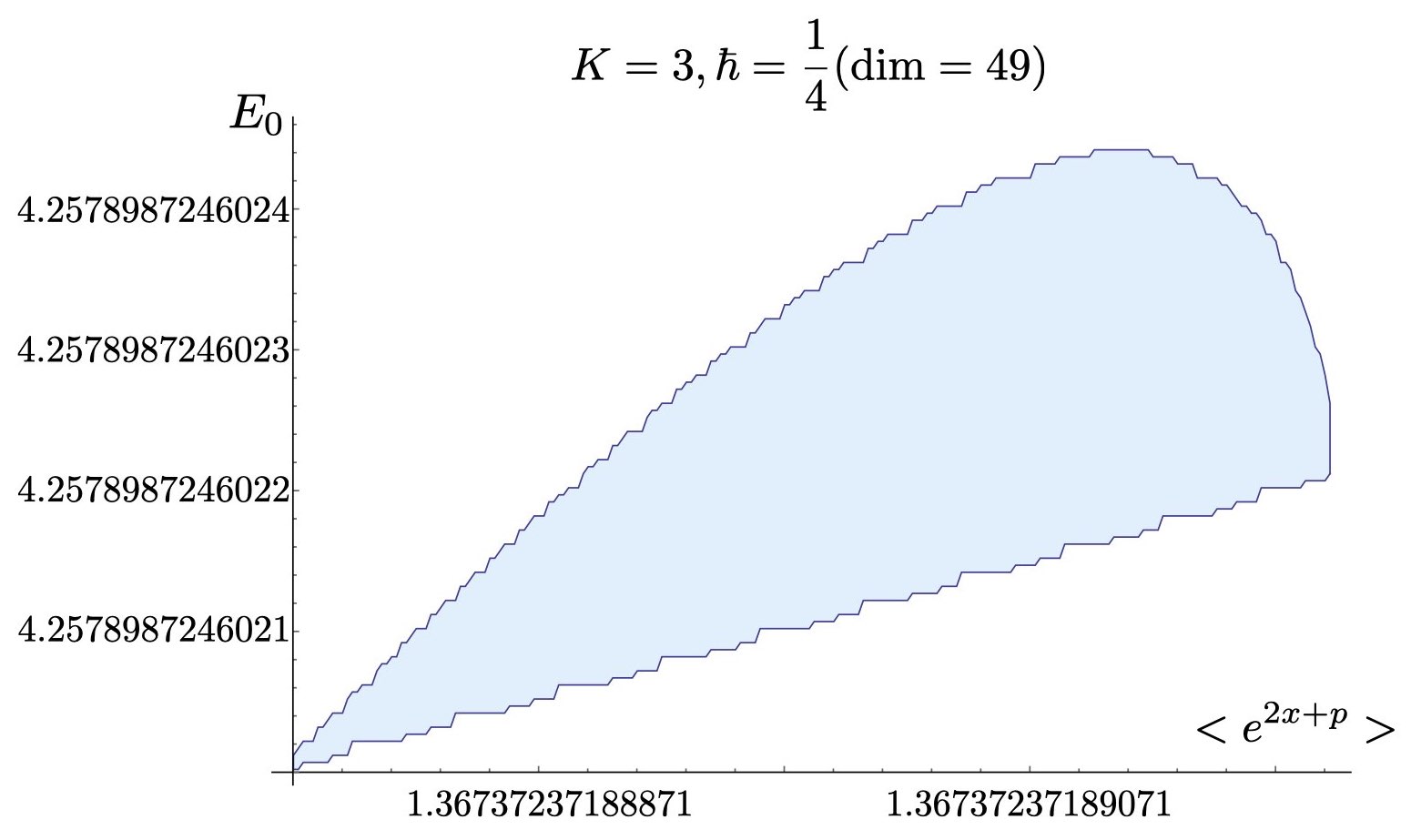}	
	\end{subfigure}
	\begin{subfigure}{0.49\textwidth}
		\includegraphics[width=\textwidth]{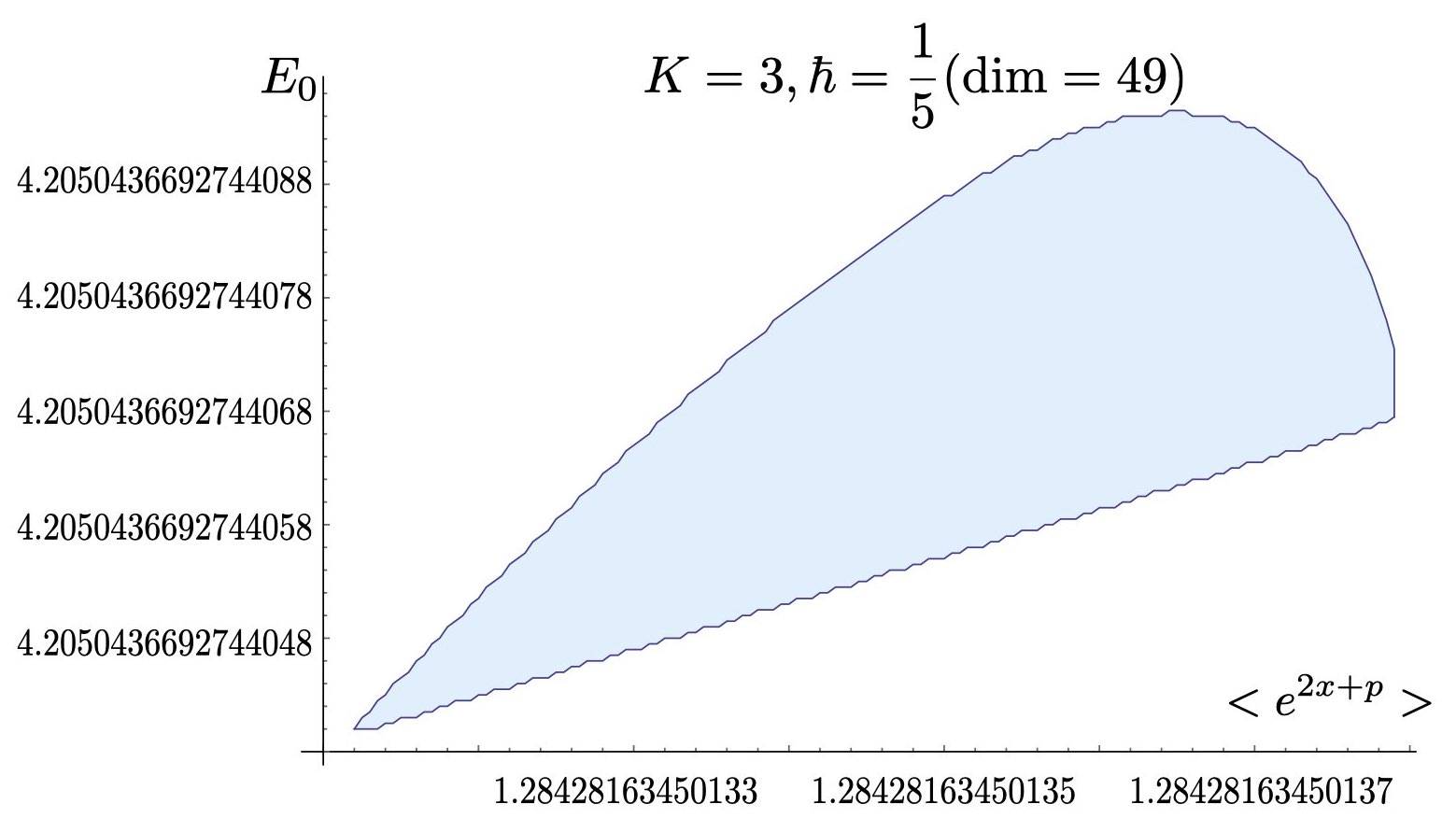}		
	\end{subfigure}
	
	 \vskip 18pt
	 
	\begin{subfigure}{0.49\textwidth}
		\includegraphics[width=\textwidth]{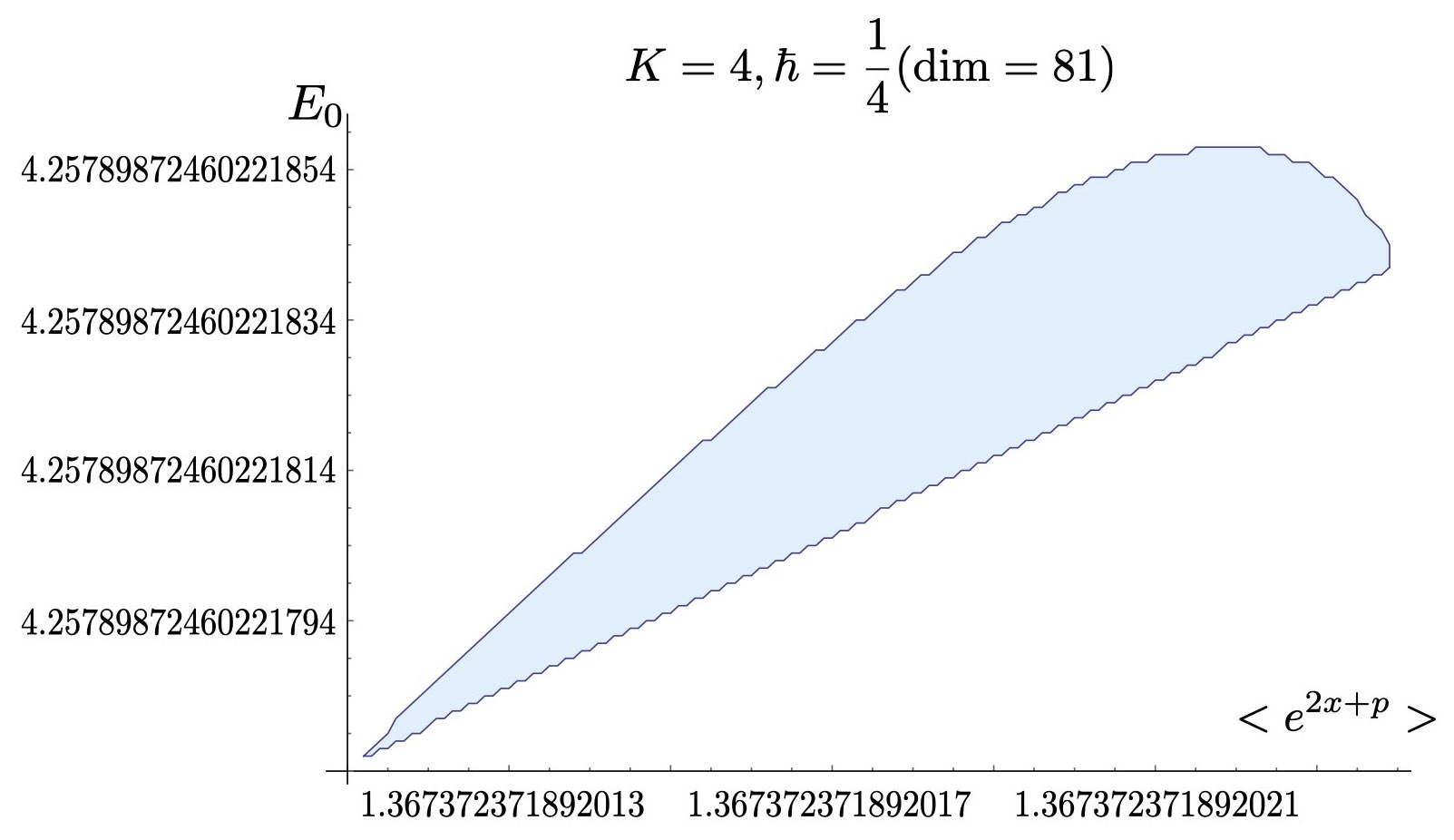}	
	\end{subfigure}
	\begin{subfigure}{0.49\textwidth}
		\includegraphics[width=\textwidth]{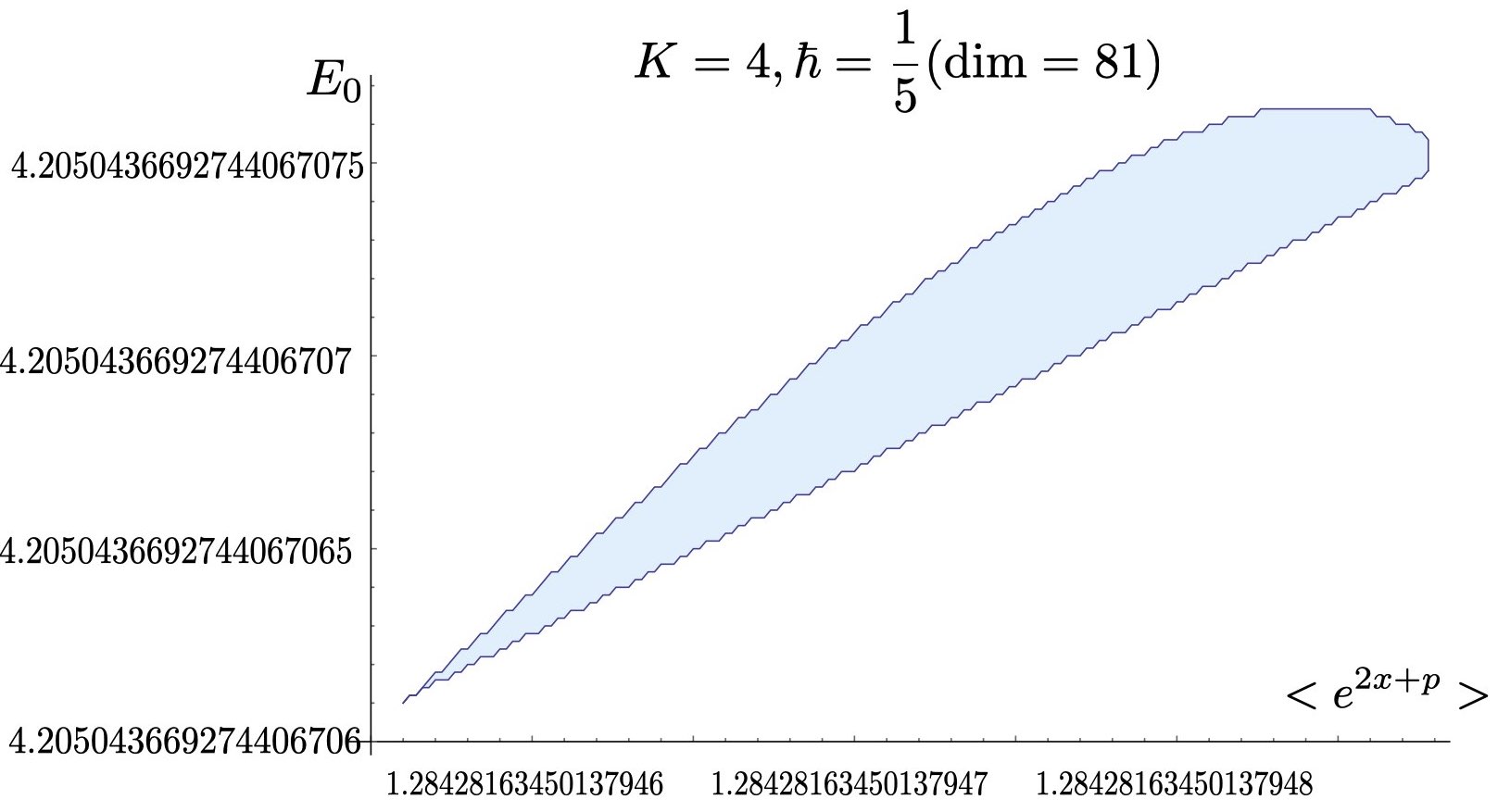}		
	\end{subfigure}
	
\caption{Bootstrap for the $\mathbb{P}^1\times \mathbb{P}^1$ model. We focus on the ground state and plot the points which satisfy the bootstrap positivity constrains for levels $K=2,3,4$. }
 \label{p1p1diagram}
\end{figure} 


\begin{table}
\begin{center}

	\begin{tabular} {|c|c|c|} \hline  Bootstrap method  & $E_0 (\hbar=\frac{1}{4})$ & $E_0 (\hbar=\frac{1}{5}) $ \\
		 \hline  $K=2$        &  \underline{4.25789872}76347275454  & \underline{4.205043669}5992931967 \\
		  \hline  $K=3$       &  \underline{4.2578987246022}426910  & \underline{4.205043669274406}8624 \\
		  \hline  $K=4$       &  \underline{4.257898724602218}2496  & \underline{4.20504366927440670}07 \\
		  \hline\hline Truncation method &4.2578987246022184123                 & 4.2050436692744067075\\
		  \hline
	\end{tabular}
	
	 \vskip 18pt
	 
	 \begin{tabular} {|c|c|c|} 
		 \hline  Bootstrap method & $\bra e^{2x+p}\ket (\hbar=\frac{1}{4})$ & $\bra e^{2x+p} \ket (\hbar=\frac{1}{5}) $ \\
		 \hline  $K=2$         &  \underline{1.3673723}626433094484   & \underline{1.28428163}36381420513 \\
		  \hline  $K=3$        &  \underline{1.36737237189}02376185   & \underline{1.2842816345013}506340\\
		  \hline  $K=4$        &  \underline{1.3673723718920}185398   & \underline{1.2842816345013794}351 \\
		  \hline \hline Truncation method &1.3673723718920241443                  & 1.2842816345013794896\\
		  \hline
	 \end{tabular}

	\caption{The estimated values of the ground state energy $E_0$ and $f_{2,1}$ for the $\mathbb{P}^1\times \mathbb{P}^1$ model. We average all the points in Fig. \ref{p1p1diagram} to get the estimated values for $\hbar=\frac{1}{4}$ and $\hbar=\frac{1}{5}$, and compare them with the truncation method using the basis of the wave eigenfunction of the quantum harmonic oscillator. The digits in the truncation method are stable with a truncation size of $300 \times 300$.  We underlie the digits in the bootstrap method which agree with the truncation method. }
\label{p1p1eigenvalue}
\end{center}

\end{table}

As we increase the truncation level, the results of the calculations should converge to their exact values.  For $|k|> \frac{2\pi}{\hbar}$, we find that the expectation value of $e^{kx}$ indeed diverges as the  truncation level increases, confirming the asymptotic behavior (\ref{asymp2.13}), while for $|k|< \frac{2\pi}{\hbar}$, except for the borderline cases, the expectation value of $e^{kx}$ has good convergence with our commonly available computational power. By the symmetry property, the expectation value of $e^{m\hat{x}+n\hat{p}}$ is finite if  $\max (|m|, |n|)< \frac{2\pi}{\hbar}$.  As a check of the formalism, we compute some convergent cases of $f_{m,n}$ and find that within the numerical accuracy, they agree with those computed from the recursion relations (\ref{recursion}) with the corresponding initial  inputs for $E$ and $f_{2,1}$. 

So the bootstrap level is bound $K\leq \frac{\pi}{\hbar}$ for this model. Otherwise, although one can still compute all $f_{m,n}$'s from the recursions (\ref{recursion}) and obtain seemingly finite results, the bootstrap procedure may fail. For example, we check that for a case $\hbar=1, K=4$,  the bootstrap matrix (\ref{matrix2.9}) always has negative eigenvalue(s), scanning the initial conditions of $E, f_{2,1}$ near and even at their physical exact values. 
 
We consider two cases $\hbar=\frac{1}{4},\frac{1}{5}$, and use the recursion relations (\ref{recursion}) to calculate the Hermitian bootstrap matrix (\ref{matrix2.9}) up to $(m,n)=(4,4)$. For the range $|m|, |n|\leq K$, the actual size of the matrix is the much bigger, i.e. it is a  $(2K+1)^2\times (2K+1)^2$ matrix. We calculate the eigenvalues of the bootstrap matrix for the levels $K=2,3,4$ and impose the positivity constrain. We focus on the ground state of the model (\ref{P1P1H}), where the approximate positions of the initial values $E_0$ and $f_{2,1}$ are known and can be also found after a rough bootstrap scan.  In Fig. \ref{p1p1diagram}, we plot the points which satisfy the bootstrap positivity constrains for levels $K=2,3,4$.  In Table \ref{p1p1eigenvalue}, we compare the results of the bootstrap method with the truncation method. We see that as we increase the level,  the positivity constrain becomes stronger and the bootstrap method achieves increasing precision.


\section{The $\mathbb{P}^2$ model} 
\label{P2}

 \begin{figure}
\centering
	\begin{subfigure}{0.49\textwidth}
		\includegraphics[width=\textwidth]{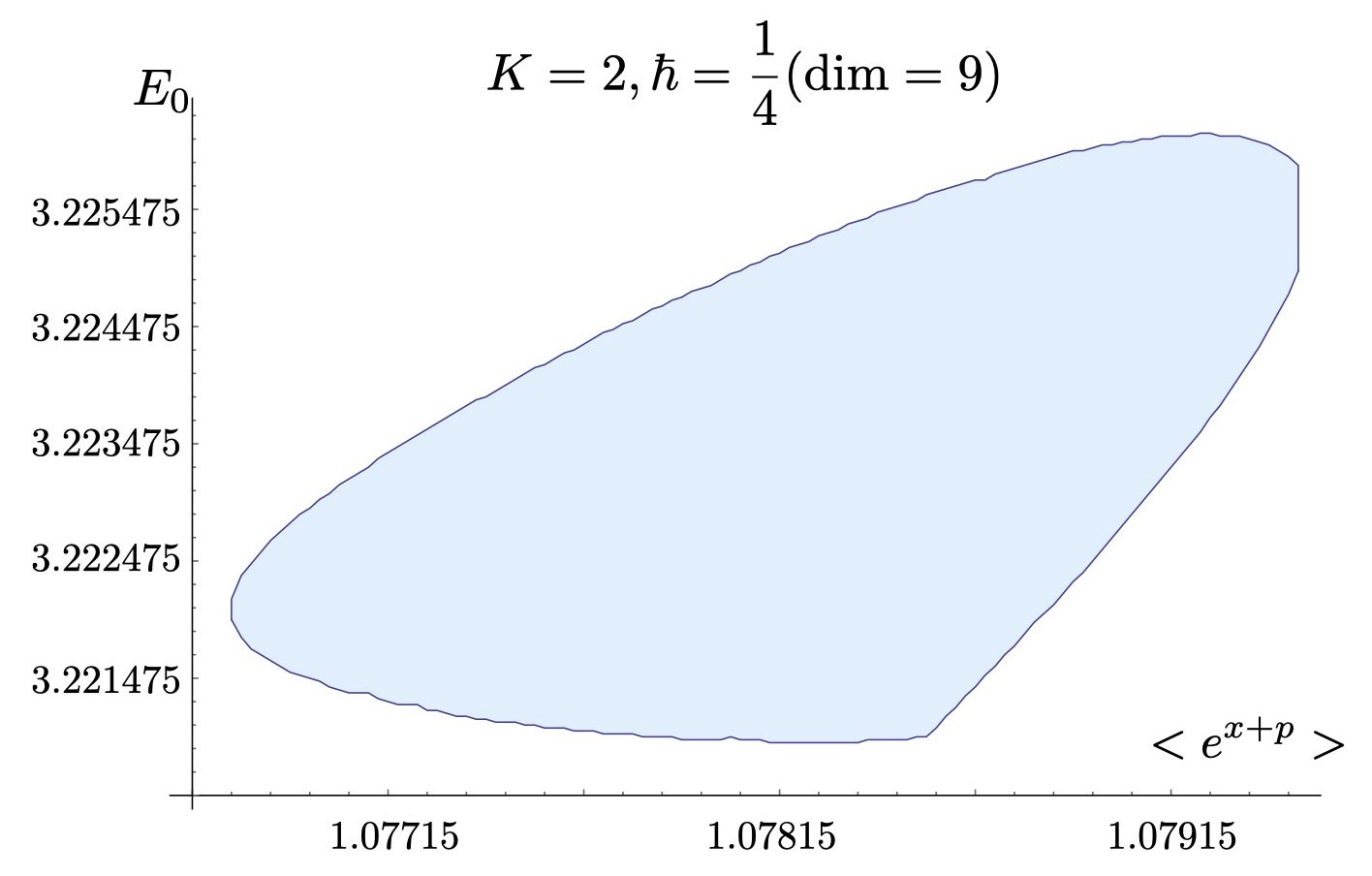} 		
	\end{subfigure}
	\begin{subfigure}{0.49\textwidth}
		\includegraphics[width=\textwidth]{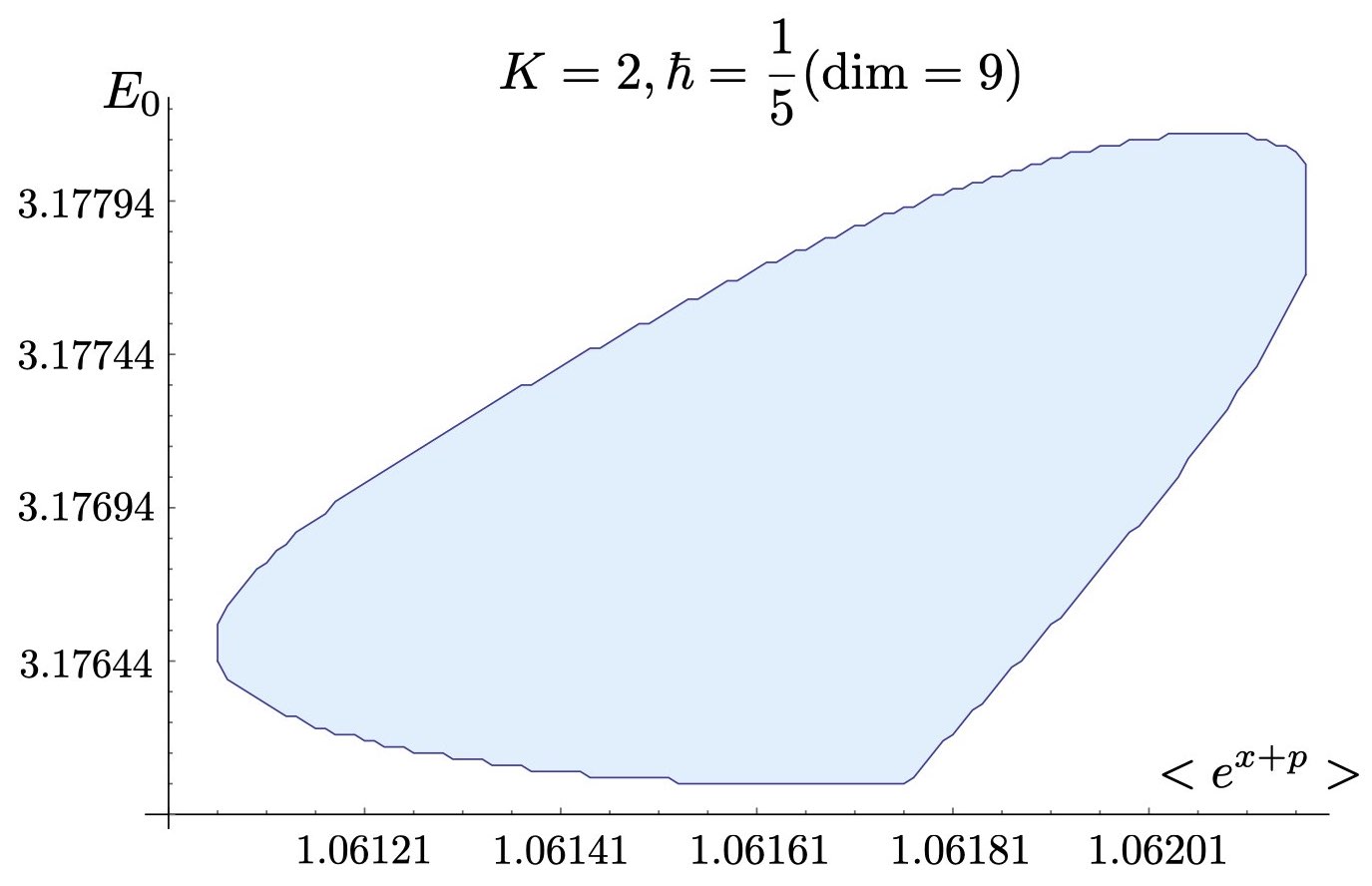} 		
	\end{subfigure}

	 \vskip 18pt

	\begin{subfigure}{0.49\textwidth}
		\includegraphics[width=\textwidth]{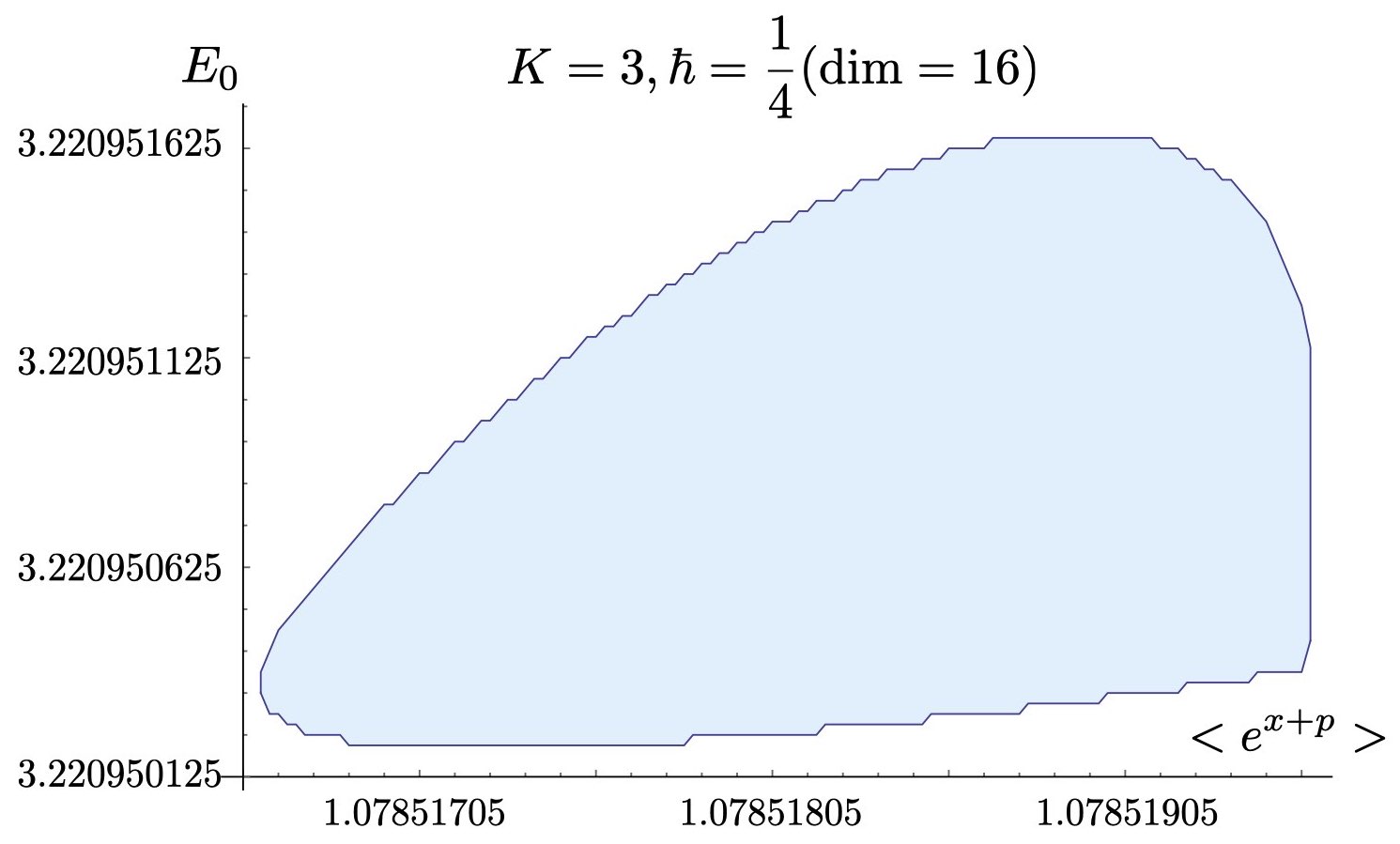}		
	\end{subfigure}
	\begin{subfigure}{0.49\textwidth}
		\includegraphics[width=\textwidth]{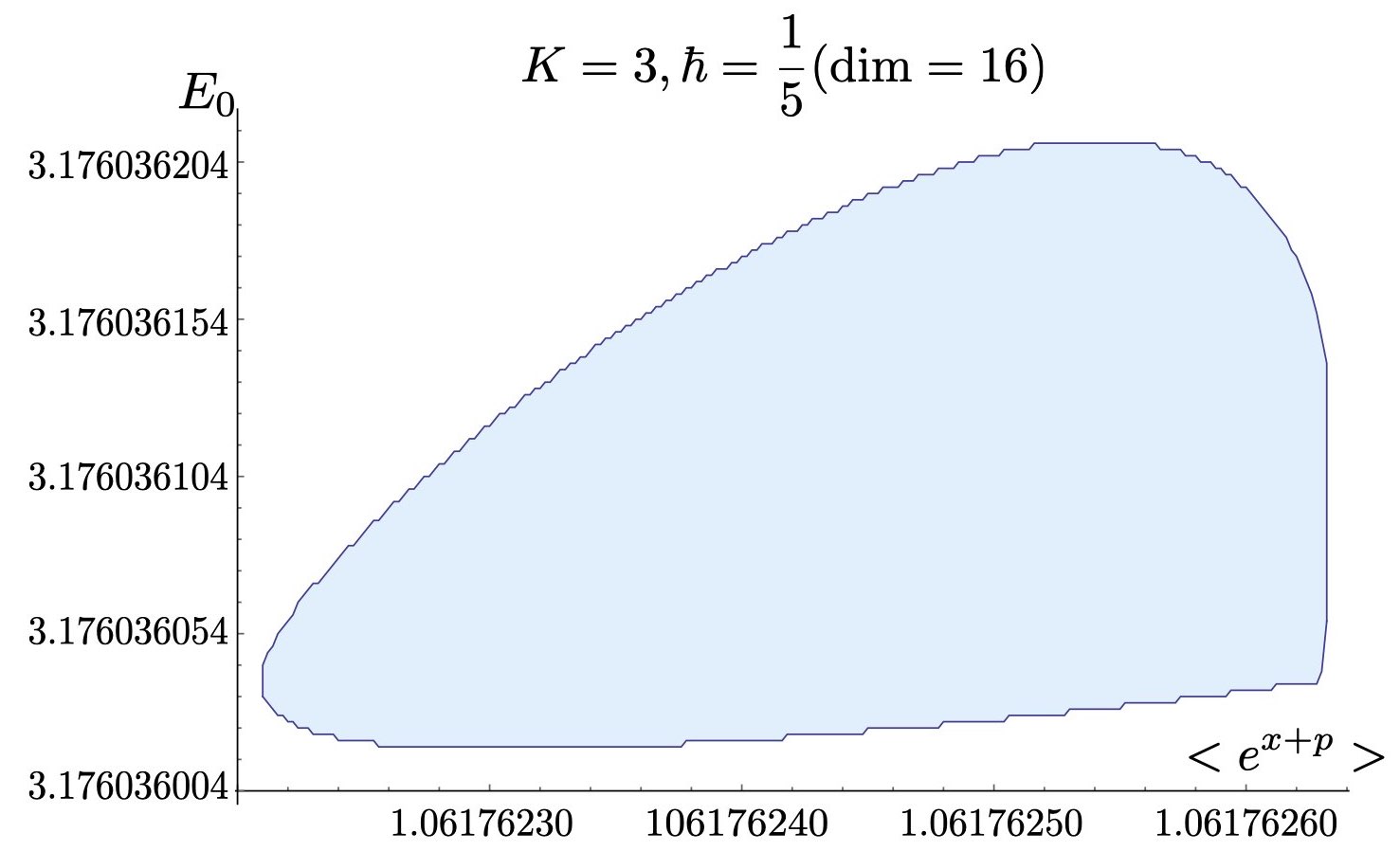}		
	\end{subfigure}
	
	 \vskip 18pt
	
	\begin{subfigure}{0.49\textwidth}
		\includegraphics[width=\textwidth]{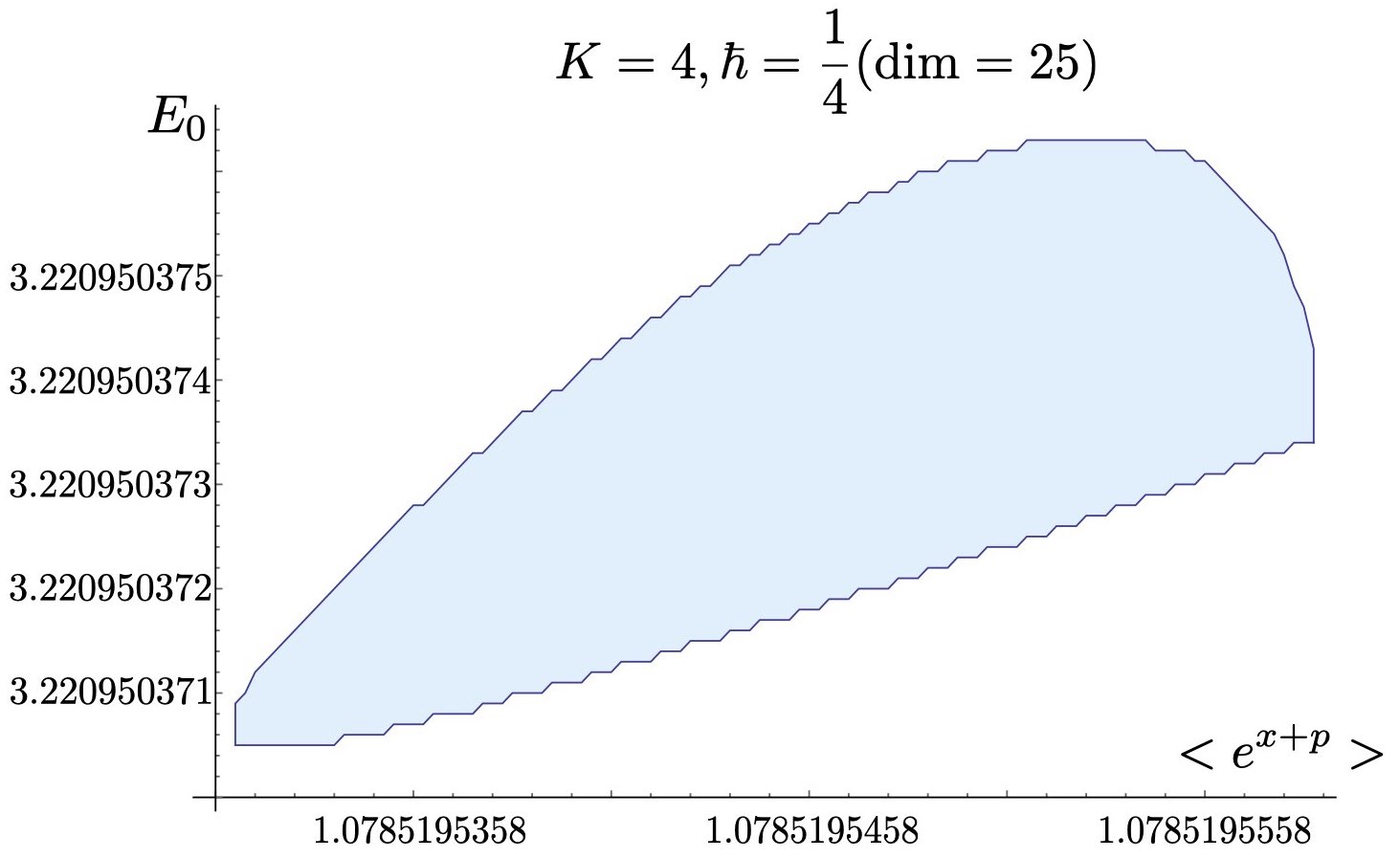}		
	\end{subfigure}
	\begin{subfigure}{0.49\textwidth}
		\includegraphics[width=\textwidth]{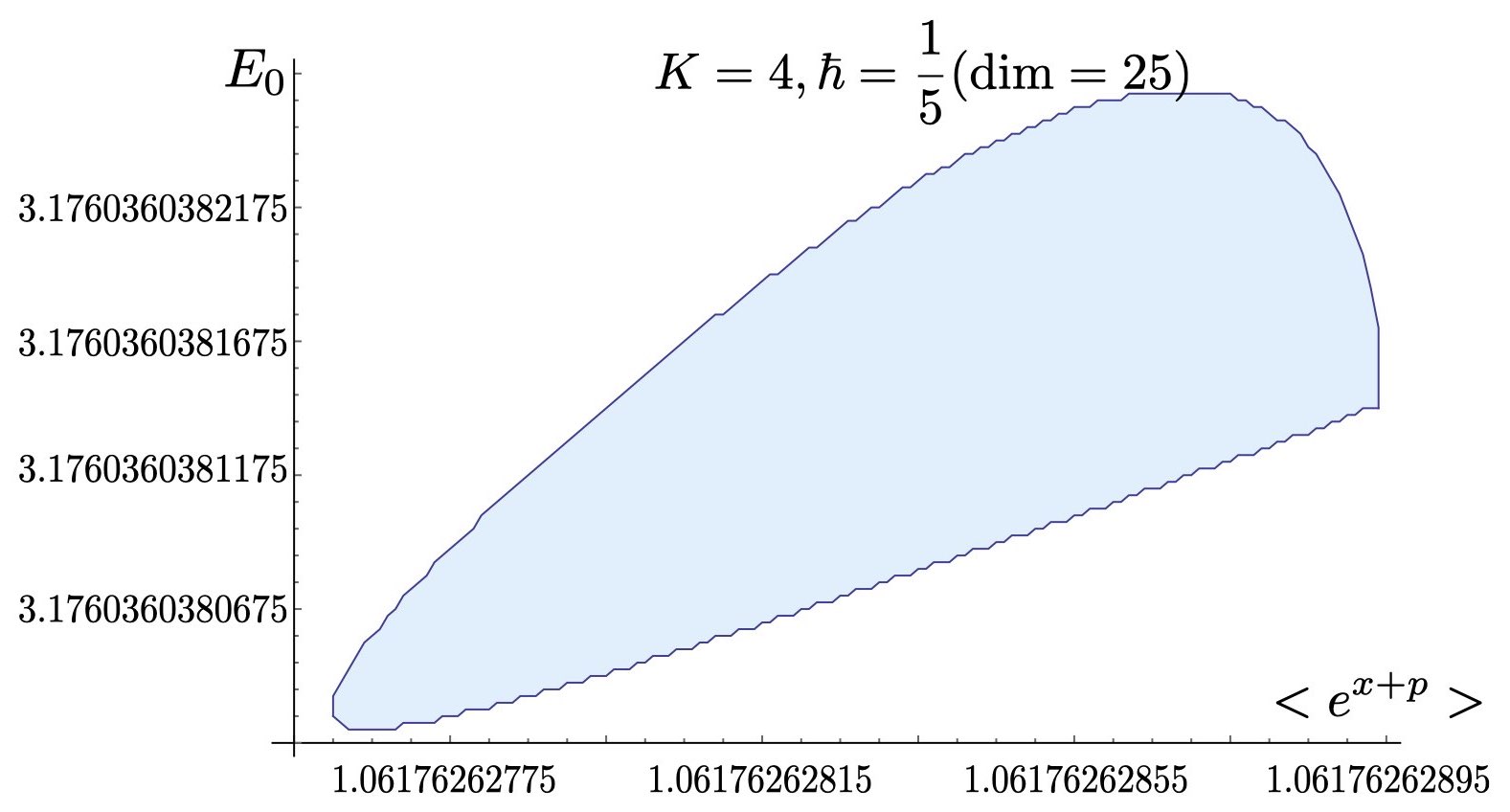}		
	\end{subfigure}
	
\caption{Bootstrap for the $\mathbb{P}^2$ model. We focus on the ground state and plot the points which satisfy the bootstrap positivity constrains for levels $K=2,3,4$. }
 \label{p2diagram}
\end{figure}

In this section, we consider the local $\mathbb{P}^2$ model, which is another simple toric Calabi-Yau geometry. The corresponding  Hamiltonian is 
\be
	\hat{H}=e^{\hat{x}} + e^{\hat{p}} + e^{-\hat{x}-\hat{p}}.
\ee
Here we also consider the expectation values of the operators $\mathcal{O}=e^{m\hat{x}+n\hat{p}}$ as before, and use the same notation 
\be
	f_{m,n}:=\bra  e^{m\hat{x}+n\hat{p}} \ket ~~~\text{and}~~~ f_{0,0}=1. 
\ee
The Hamiltonian is invariant under the transformation: 
\be
	(x,p)\rightarrow (p,x)~~~~\text{and}~~~\hbar \rightarrow -\hbar,
\ee
so similarly as before, we have the following symmetry 
\be \label{p2 symmetry}
	f_{m,n} = f_{n,m}.
\ee

\begin{table}
\begin{center}

	\begin{tabular} {|c|c|c|} \hline  Bootstrap method  & $E_0 (\hbar=\frac{1}{4})$ & $E_0 (\hbar=\frac{1}{5}) $ \\
		 \hline  $K=2$           &  \underline{3.22}32954229458538432 & \underline{3.17}70057998600419874 \\
		  \hline  $K=3$          &  \underline{3.220950}8284518416388 & \underline{3.1760360}966894729711 \\
		  \hline  $K=4$          &  \underline{3.2209503734}779853031 & \underline{3.17603603814}46034538\\
		  \hline \hline  Truncation method &3.2209503734162626526                  &3.1760360381435672645\\
		  \hline
	\end{tabular}
	
	 \vskip 18pt
	 
	 \begin{tabular} {|c|c|c|} 
		 \hline  Bootstrap method & $\bra e^{x+p}\ket (\hbar=\frac{1}{4})$ & $\bra e^{x+p} \ket (\hbar=\frac{1}{5}) $ \\
		 \hline  $K=2$          &  \underline{1.078}2066679288148429  & \underline{1.061}6524310706787964 \\
		  \hline  $K=3$         &  \underline{1.07851}83017569430766  & \underline{1.061762}4481052702886 \\
		  \hline  $K=4$         &  \underline{1.0785195}464312002449  & \underline{1.061762628}3451065203\\
		  \hline \hline  Truncation method &1.0785195590187190830                  &1.0617626289497548351\\
		  \hline
	 \end{tabular}

	\caption{The estimated values of the ground state energy $E_0$ and $f_{1,1}$ for the $\mathbb{P}^2$ model. We average all the points in Fig. \ref{p2diagram} to get the estimated values for $\hbar=\frac{1}{4}$ and $\hbar=\frac{1}{5}$, and compare them with the truncation method. The digits in the truncation method are stable with a truncation size of $300 \times 300$.  We underlie the digits in the bootstrap method which agree with the truncation method. }

\label{p2eigenvalue}
\end{center}

\end{table}

We use the same bootstrap relations (\ref{bootstrap relations}), which in this  case give the similar recursion relations 
\be \ba  
\sin(\frac{n\hbar}{2}) f_{m+1,n}  - \sin(\frac{ m \hbar}{2}) f_{m,n+1} + \sin(\frac{(m-n)\hbar}{2}) f_{m-1,n-1}&=  0, \\
\cos(\frac{n\hbar}{2}) f_{m+1,n}  + \cos(\frac{ m \hbar}{2})  f_{m,n+1} + \cos(\frac{(m-n)\hbar}{2}) f_{m-1,n-1}  &= Ef_{m,n}.
\ea  \ee

In local $\mathbb{P}^2$ model, because of the symmetry (\ref{p2 symmetry}), we have $E=3f_{1,0}$. We also need two initial conditions, chosen to be the energy $E$ and $f_{1,1}$,  for the recursion relations.

The initial conditions $E$ and $f_{1,1}$ can determine all $f_{m,n}$'s with $m,n\geq 0$ from the recursion relations. But unlike the previous case,  the  $\mathbb{P}^2$ model does not have the symmetry $x \rightarrow -x$ or $ p \rightarrow -p $, so there is no symmetry for switching the signs of indices in $f_{m,n}$.  The calculations of these $f_{m,n}$'s with $m<0$ or $n<0$ would need a different initial condition. Thus to minimize the number of initial conditions in the bootstrap procedure, we focus only on $f_{m,n}$'s with $m,n\geq 0$ in the similar operator (\ref{operator2.8}) for the $\mathbb{P}^2$  model.  As a result, at level $K$ the bootstrap matrix  (\ref{matrix2.9}) is a smaller $(K+1)^2\times (K+1)^2$ matrix.


The asymptotic behavior of the normalizable wave function is the same as the previous $\mathbb{P}^1\times \mathbb{P}^1$ model in (\ref{asymp2.13}) by a similar WKB analysis. So for a Planck constant $\hbar$, the available operators $f_{m,n}$'s for bootstrap  are in the range $0\leq m,n \leq \frac{2\pi}{\hbar}$.

We perform the similar analysis as in the previous case, also consider the cases of $\hbar=\frac{1}{4},\frac{1}{5}$ and calculate the bootstrap matrix (\ref{matrix2.9}) up to $(m,n)=(4,4)$,  focusing on the ground state.  In Fig. \ref{p2diagram}, we plot the points which satisfy the bootstrap positivity constrains for levels $K=2,3,4$.  In Table \ref{p2eigenvalue}, we compare the results of the bootstrap method with the truncation method. Due to the smaller size of the bootstrap matrix, the numerical precision of the $\mathbb{P}^2$ model is lower than that of the previous $\mathbb{P}^1\times \mathbb{P}^1$ model. For example, the precision of the $K=4$ level in the $\mathbb{P}^2$ model is comparable that of $K=2$ level in the $\mathbb{P}^1\times\mathbb{P}^1$ model.


\section{A non-relativistic Toda model} 
\label{SU2} 

In this section we consider a case of different type of models, obtained e.g. from the class of quantum systems in \cite{Grassi:2018bci} by a simple exchange of $\hat{x}$ and $\hat{p}$ operators.  The Hamiltonian is 
\be  \label{HSU2}
\hat{H} = \frac{\hat{p}^2}{2} + \cosh(\hat{x}) . 
\ee
This also belongs to the class of non-relativistic Toda integrable models, and has been long studied in the literature, see e.g. the  papers \cite{Nekrasov:2009rc, Kozlowski:2010tv} in the context of Nekrasov-Shatashvili quantization conditions and references therein. Our model is basically equivalent to the simplest two-body case of the Toda chain models. 

The exact quantization condition is first derived in \cite{Gutzwiller:1980yx}, and in the modern approach is given by the pure $SU(2)$ Seiberg-Witten theory.  We note that although the potential is related to the periodic cosine function by a simple rotation $x\rightarrow ix$ and the perturbative WKB calculations are also simply related, the underlying physics of this model is actually quite different from the one considered in \cite{Berenstein:2021loy}. In particular, the system have bound states with quantized energies and normalizable wave functions over real $x$, instead of the periodic wave functions from the well known Mathieu's differential equation. Unlike the Calabi-Yau models in the previous sections or conventional non-relativistic quantum mechanics with polynomial potentials, the quantization of the model (\ref{HSU2}) using Nekrasov partition function is perturbative in $\hbar$, without non-perturbative contributions of the non-analytics form $e^{-\frac{A}{\hbar}}$.

The one-dimensional operators for bootstrap are similar to \cite{Berenstein:2021loy} by a rotation $x\rightarrow ix$. As in the Calabi-Yau models, we consider a more general class of two-dimensional operators and denote the expectation values $f_{m,n}:= \bra \hat{p}^m e^{n\hat{x}} \ket$,  which may now be complex since the operators are not necessarily Hermitian. With the proper normalization of the energy eigenstate we have  $f_{0,0}=1$. Since the potential is an even function, we can choose the energy eigenfunctions to be either even or odd functions. Therefore there is a symmetry property $f_{m,n}=(-1)^mf_{m,-n}$, where we only consider non-negative power of momentum operator with the integer $m\geq 0$.

The relevant bootstrap equations are 
\be 
 \bra \hat{H} \hat{p}^m e^{n\hat{x}} \ket  =   \bra \hat{p}^m e^{n\hat{x}}  \hat{H}  \ket  = E \bra \hat{p}^m e^{n\hat{x}} \ket . 
\ee
Using the formula $e^{\hat{x}} \hat{p}^n = ( \hat{p}+ i\hbar)^n  e^{\hat{x}}$, we obtain the recursion relations among $f_{m,n}$'s 
\be \ba \label{recur4.2}
&  ~~~~ \bra \hat{p}^{m+2} e^{n\hat{x}}   +(\hat{p}+i\hbar)^{m} e^{(n+1)\hat{x}}   +(\hat{p} - i\hbar)^{m} e^{(n-1)\hat{x}}    \ket  \\
&=   \bra \hat{p}^{m}  (\hat{p}+in \hbar)^{2} e^{n\hat{x}}   +\hat{p}^{m} (e^{(n+1)\hat{x}}   + e^{(n-1)\hat{x}}  )  \ket  \\
&= 2E \bra \hat{p}^m e^{n\hat{x}} \ket 
\ea \ee
One can also eliminate the $\hat{p}$ operator from the equations to have a recursion relation on the sub-class of one-dimensional operators 
\be  \label{recur4.3}
n(4E+\frac{ n^2 \hbar^2}{2}) f_{0,n} = (2n+1)f_{0, n+1} +(2n-1) f_{0, n-1} . 
\ee
Similarly to the toric Calabi-Yau geometries, we also need two initial conditions for the recursive computations, which are chosen as $E$ and $f_{0,1}$. One can first solve for all $f_{0,n}$'s with the recursion (\ref{recur4.3}), then the general $f_{m,n}$ can be determined from (\ref{recur4.2}) and depends on $f_{0,n+k}$'s for $|k|\leq [\frac{m}{2}]$.

In order to understand the appropriate ranges of the indices for bootstrap, we need to analyze the asymptotic behavior of the energy eigenfunctions. The leading order WKB equation of the wave function gives two branches 
\be
\psi(x) \sim \exp(\pm\frac{2e^{\frac{|x|}{2}}}{\hbar}),~~~ x\rightarrow \pm \infty.
\ee
 As familiar in the theory of linear differential equations, and also analyzed in details in \cite{Gutzwiller:1980yx} in this case, for generic values of energy $E$, the divergent branch of wave function can not be cancelled at both $x\rightarrow \pm \infty$ for any linear combination of the two independent solutions of the Schr\"{o}dinger equation. For the physical energy satisfying the quantization condition, it is possible to find a  linear combination which cancels the divergence at both $x\rightarrow \pm \infty$, so that the resulting wave function may be square integrable. The decaying branch in the WKB analysis is always present and usually provides the actual asymptotic behavior of the normalizable eigenfunction, as it happens in the cases of the Calabi-Yau models in previous sections and also e.g. the harmonic oscillator. Although this is not so clear from the wave function constructed in \cite{Gutzwiller:1980yx}, it is likely that this is also true for this model, which would ensure the finiteness of $f_{0,n}$'s for all $n\in \mathbb{Z}$ and their availability for bootstrap.  It would be interesting to investigate this issue further with some modern analytic approaches.  Here for our purpose, we check numerically with the truncation method that the expectation values $f_{0,n}$ are indeed always finite.  The momentum operator can also modify the asymptotic behavior. For example, if there is an oscillatory factor $\exp(i e^x)$ in the wave function, the action of the momentum operator $\hat{p}$ would produce a divergent factor $e^x$ as $x\rightarrow +\infty$. In our case, since the general $f_{m,n}$'s are determined in terms of $f_{0,n}$'s, they should be finite as well. Another perspective is to use the wave function in momentum space. By a similar WKB analysis as in the Calabi-Yau models, one easily finds the same asymptotic behavior $\psi(p)\sim e^{-\frac{\pi}{\hbar} |p|}$ as the momentum $p\rightarrow \pm \infty$, so the expectation value of $\hat{p}^m$ for any $m\geq 0$ is finite.

We provide some technical details for checking the finiteness of $f_{0,n}$ with the truncation method. In principle one could use the eigenfunctions of a harmonic oscillator with arbitrary mass and frequency for truncation. However, as in \cite{Huang:2014eha}, in practice the calculations work better with appropriate empirical choices of mass/frequency. For example, consider $\hbar=1$ and use a harmonic oscillator with unit mass and frequency $\omega$ for truncation. We find that in this case, the range of $\omega\in (2,10)$ is best for the calculations. As we increase the truncation level, the expectation values  $f_{0,n}$ quickly converge and agree with the results from the recursion relation (\ref{recur4.3}). For smaller $\omega$ and large $n$, e.g. $\omega=1$ and $n>7$, there appears to be a false divergence of  $f_{0,n}$ as we increase the truncation level. Although we believe the calculations should eventually converge, it is beyond our computational ability to check this point explicitly. For larger  $\omega$, the computations converge more slowly, with no (false) appearance of divergence. On the other hand, we also check that the true divergence in the Calabi-Yau models in the previous sections can not be eliminated by such choices of the frequency.


It turns out that the bootstrap method with just the one indexed $f_{0,n}$'s  does not provide a good constrain for the energy eigenvalue for this model. It is necessary to use the two-indexed expectation value $f_{m,n}$'s and consider the operator  
\be 
	\mathbb{O} = \sum_{m,n} c_{m,n} \hat{p}^m e^{n \hat{x} }. 
\ee
Similar to previous models, the positivity of the $\bra \mathbb{O} \mathbb{O} ^{\dagger} \ket $ is equivalent to positivity  of the Hermitian bootstrap matrix 
\be \label{matrixToda}
	M_{(m,n),(m',n')}=\bra \hat{p}^m (p+ i (n+n^\prime ) \hbar)^{m'} e^{(n+n')x} \ket, 
\ee 
whose matrix elements are simply linear combinations of some two-indexed expectation values by expanding out the $m^\prime$ power.

\begin{figure}
\centering
	\begin{subfigure}{0.49\textwidth}
		\includegraphics[width=\textwidth]{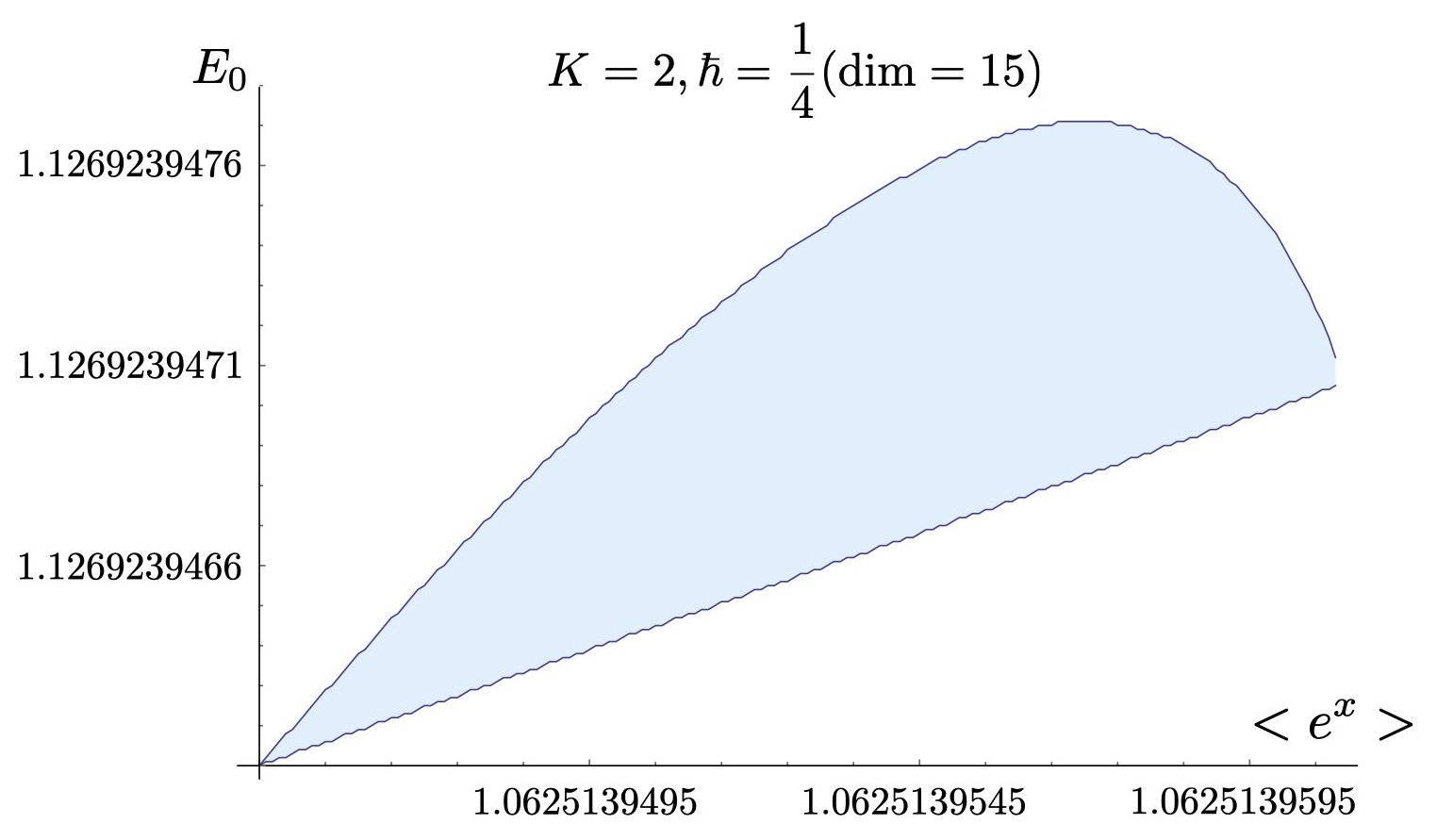} 		
	\end{subfigure}
	\begin{subfigure}{0.49\textwidth}
		\includegraphics[width=\textwidth]{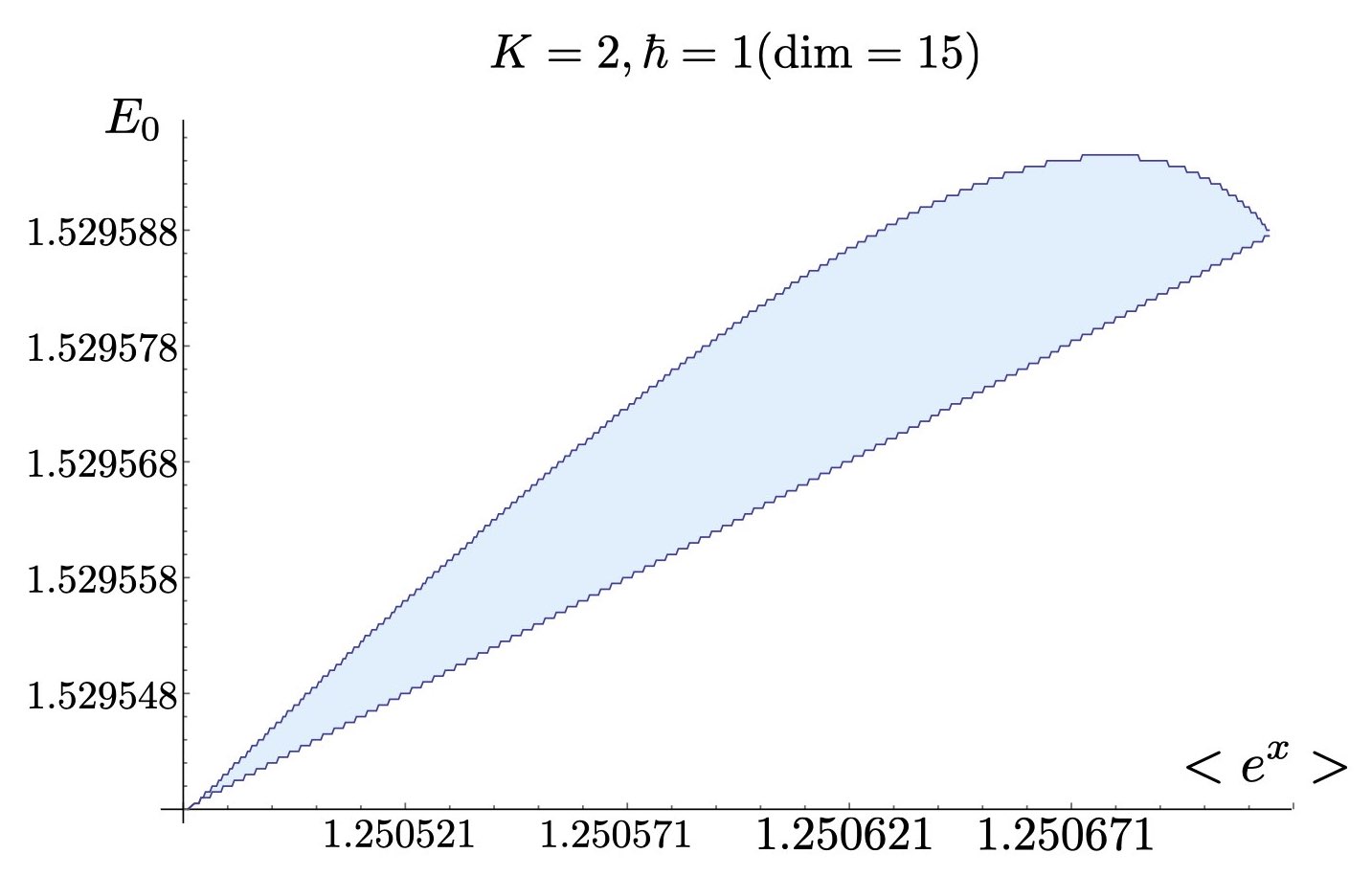} 		
	\end{subfigure}

	 \vskip 18pt

	\begin{subfigure}{0.49\textwidth}
		\includegraphics[width=\textwidth]{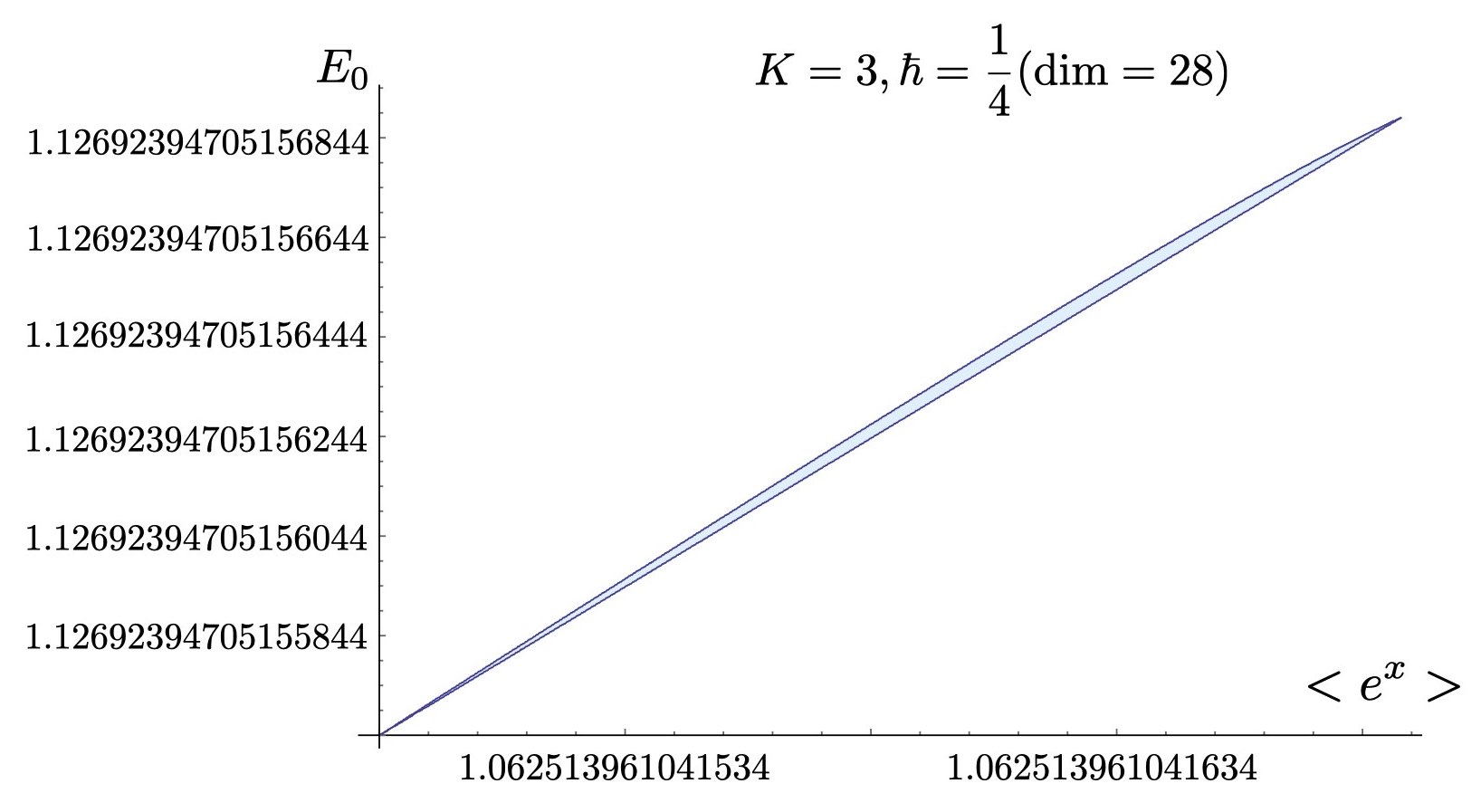}		
	\end{subfigure}
	\begin{subfigure}{0.49\textwidth}
		\includegraphics[width=\textwidth]{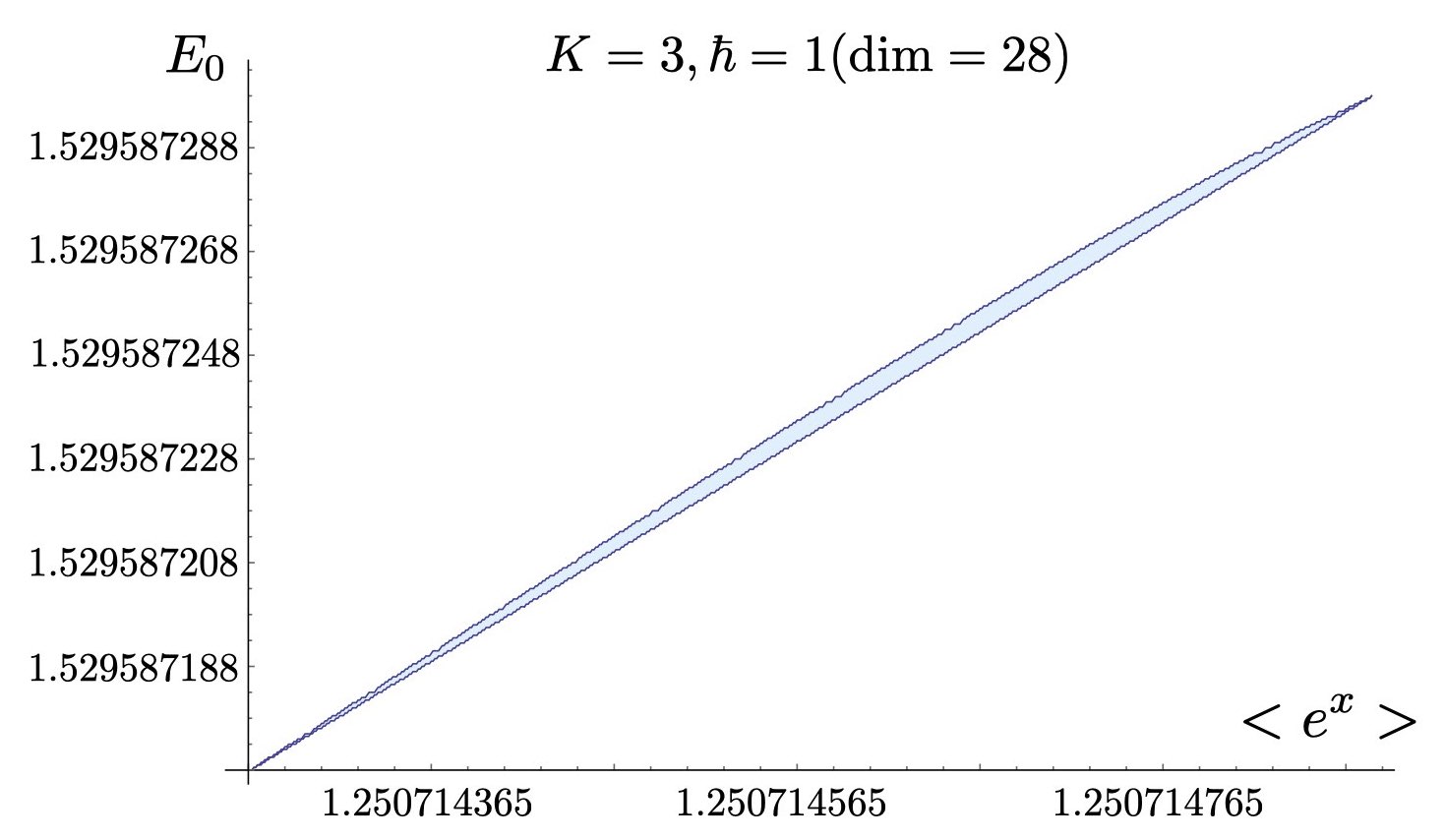}		
	\end{subfigure}
	
	 \vskip 18pt
		
\caption{ Bootstrap for the two-body non-relativistic Toda model. We focus on the ground state and plot the points which satisfy the bootstrap positivity constrains for levels $K=2,3$. }
 \label{mathieudiagram}
\end{figure}

We perform the similar analysis as in the previous cases, consider two cases of $\hbar=1, \frac{1}{4}$ and calculate the bootstrap matrix (\ref{matrixToda}) up to level $(m,n)=(3,3)$, focusing also on the ground state.  In Fig. \ref{mathieudiagram}, we plot the points which satisfy the bootstrap positivity constrains for levels $K=2,3$.  In Table \ref{mathieueigenvalue}, we compare the results of the bootstrap method with the truncation method. It turns out that the bootstrap works much better for this model than the Calabi-Yau models in the previous sections.  We are able to compute for a larger range of the Planck constant, e.g. for the case $\hbar=1$, it would be difficult to achieve a meaningful numerical accuracy with the bootstrap method in the Calabi-Yau models. While for the case $\hbar=\frac{1}{4}$, although the size of the bootstrap matrix is much smaller, the precision of the $K=3$ level in the non-relativistic Toda model is already comparable that of $K=4$ level in the $\mathbb{P}^1\times\mathbb{P}^1$ model.  

\begin{table}
\begin{center}

	\begin{tabular} {|c|c|c|} \hline  Bootstrap method  & $E_0 (\hbar=\frac{1}{4})$ & $E_0 (\hbar=1) $ \\
		  \hline  $K=2$          &  \underline{1.1269239470}057694962  & \underline{1.5295}730767977848359 \\
		  \hline  $K=3$          &  \underline{1.12692394705156}35115  & \underline{1.5295872}406198638463 \\
		  \hline \hline  Truncation method  &1.1269239470515689056                  &1.5295872984507625053\\
		  \hline
	\end{tabular}
	
	 \vskip 18pt
	 
	 \begin{tabular} {|c|c|c|} 
		 \hline  Bootstrap method & $\bra e^{x}\ket (\hbar=\frac{1}{4})$ & $\bra e^{x} \ket (\hbar=1) $ \\
		 \hline   $K=2$          & \underline{1.0625139}538919156068  & \underline{1.250}6087617069793957 \\
		  \hline  $K=3$          & \underline{1.0625139610416}004653  & \underline{1.250714}6011718939947 \\
		  \hline \hline  Truncation method &1.0625139610416930424                  &1.2507148830512434453\\
		  \hline
	 \end{tabular}
	
	\caption{	The estimated values of the ground state energy $E_0$ and $f_{1,1}$ for the two-body non-relativistic Toda model. We average all the points in Fig. \ref{p2diagram} to get the estimated values for $\hbar=\frac{1}{4}$ and $\hbar=1$, and compare them with the truncation method. The digits in the truncation method are stable with a truncation size of $300 \times 300$.  We underlie the digits in the bootstrap method which agree with the truncation method.  }

\label{mathieueigenvalue}
\end{center}
\end{table}


\section{The quartic anharmonic oscillator}
\label{secquartic}

Motivated by our studies in the previous sections, we apply our improved bootstrap method which is recursive in both position and momentum operators to the quantum quartic  anharmonic oscillator, which have been studied in the recent bootstrap literature \cite{Han:2020bkb, Berenstein:2021loy, Bhattacharya:2021btd}. The Hamiltonian is 
\be
	\hat{H}=\hat{p}^2+\hat{x}^2+g \hat{x}^4.
\ee
To compare our method with that of \cite{Han:2020bkb}, we consider the same choice of parameters $\hbar=1,g=1$. The case of a negative quadratic term gives a double-well potential and is considered in \cite{Berenstein:2021loy, Bhattacharya:2021btd}. The quantization of this model also has a long history, see e.g. the early works \cite{Bender:1969si, Zinn-Justin:1981qzi} as well as a more recent paper \cite{Krefl:2013bsa}  in the context of  Nekrasov-Shatashvili quantization conditions \cite{Nekrasov:2009rc} and Dijkgraaf-Vafa matrix models \cite{Dijkgraaf:2009pc}. 

We denote the complex expectation values in a properly normalized energy eigenstate similarly as $f_{m,n}\equiv \bra \hat{p}^m \hat{x}^n \ket$ with $m,n\geq 0$ . The one-dimensional recursion relation is given in \cite{Han:2020bkb} as
\be
	4nE f_{0,n-1}  + n(n-1) (n-2) f_{0,n-3}  - 4(n+1) f_{0,n+1}  -4g(n+2) f_{0,n+3} = 0,
\ee	
which can be solved recursively with the energy eigenvalue $E$ and an additional initial condition $f_{0,2}= \bra x^2 \ket$. In the one-dimensional approach, one uses the real symmetric bootstrap matrix $\mathcal{M}_{ij}=\bra x^{i+j}\ket$ for imposing the positivity constraint.

\begin{figure}

	\begin{subfigure}{0.49\textwidth}
		\includegraphics[width=\textwidth]{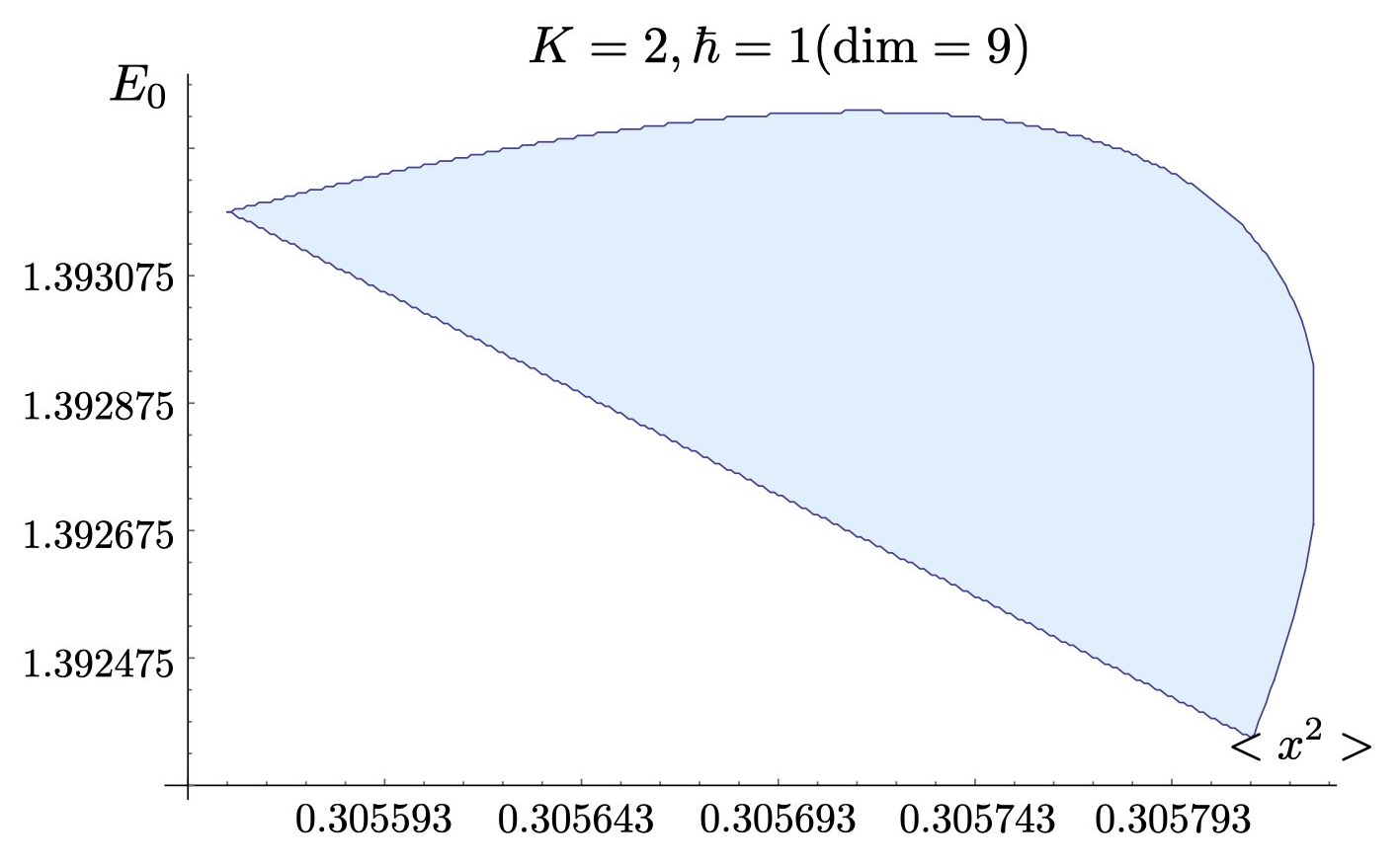}		
	\end{subfigure}
	\begin{subfigure}{0.49\textwidth}
		\includegraphics[width=\textwidth]{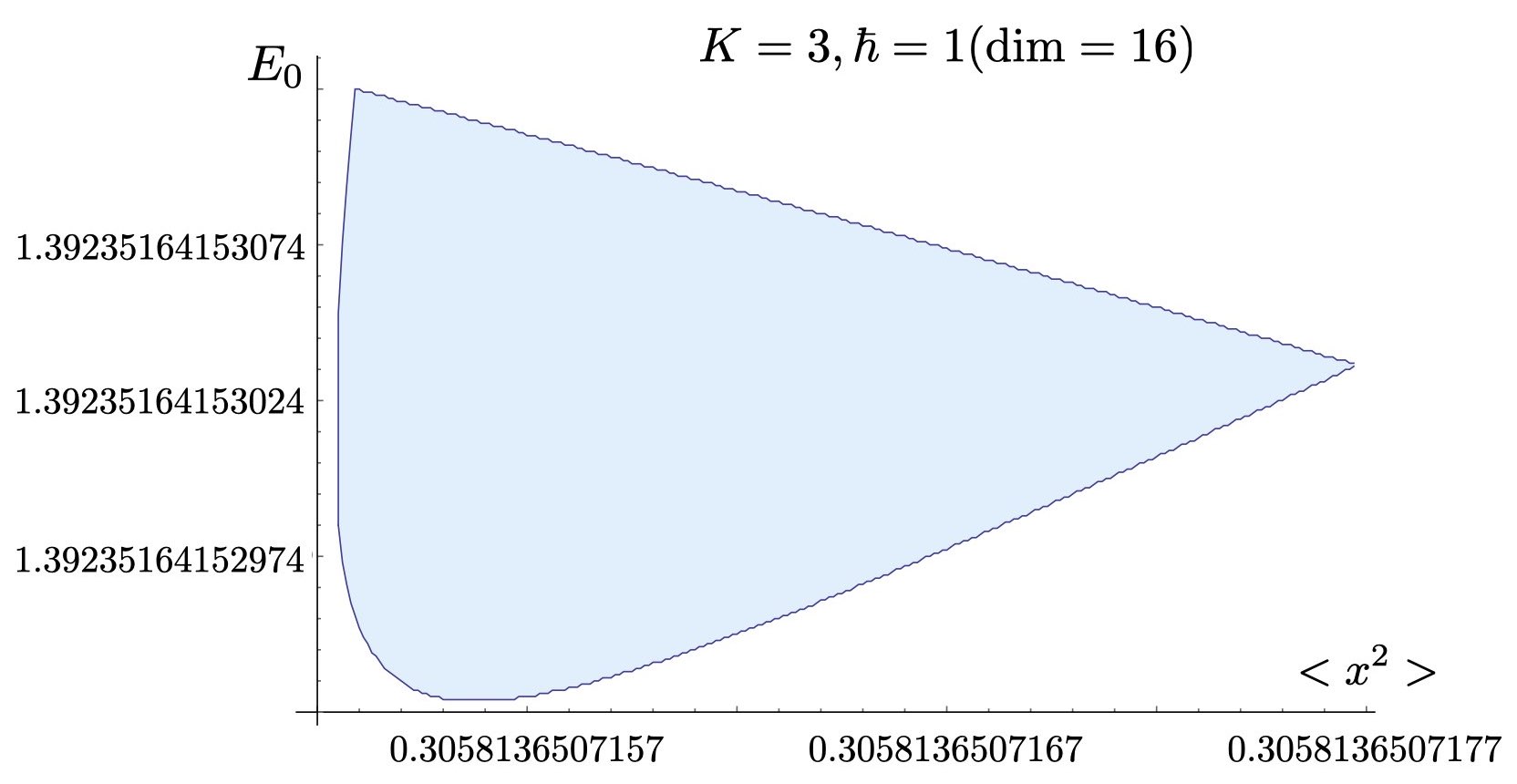}		
	\end{subfigure}
	
	 \vskip 18pt
	 
	 \begin{subfigure}{0.49\textwidth}
		\includegraphics[width=\textwidth]{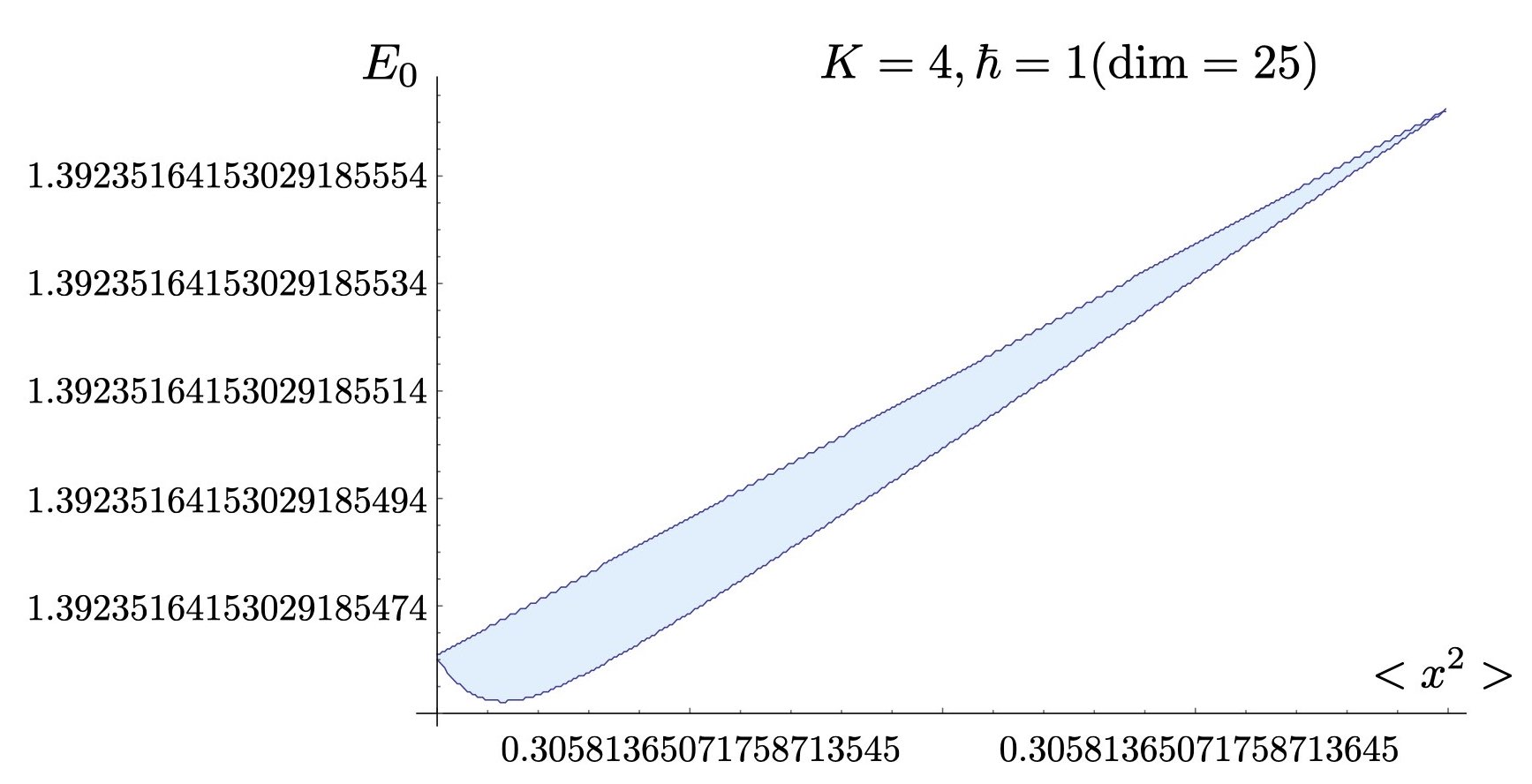} 		
	\end{subfigure}
         
          \vskip 18pt
	
	\begin{subfigure}{0.49\textwidth}
		\includegraphics[width=\textwidth]{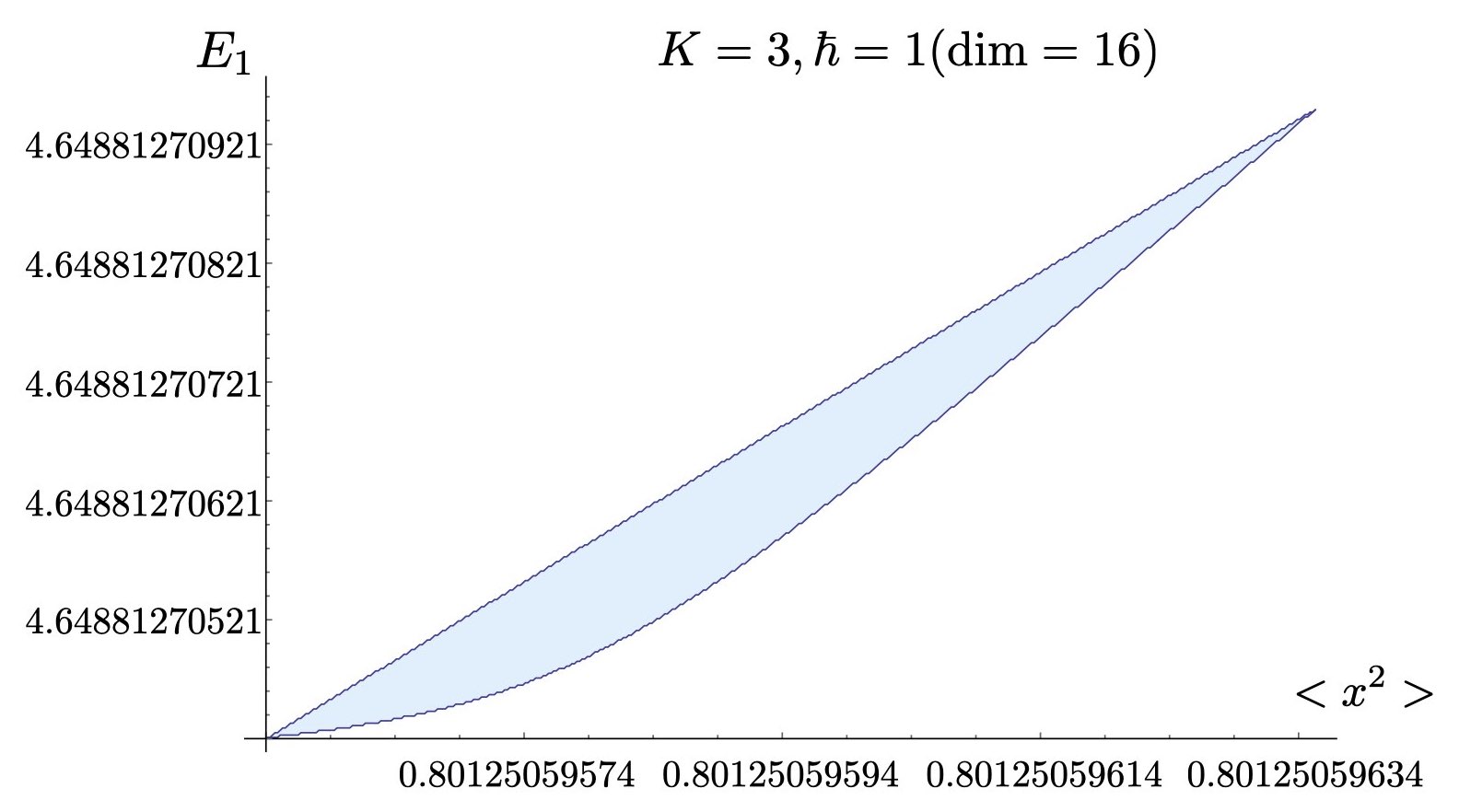}		
	\end{subfigure}
	\begin{subfigure}{0.49\textwidth}
		\includegraphics[width=\textwidth]{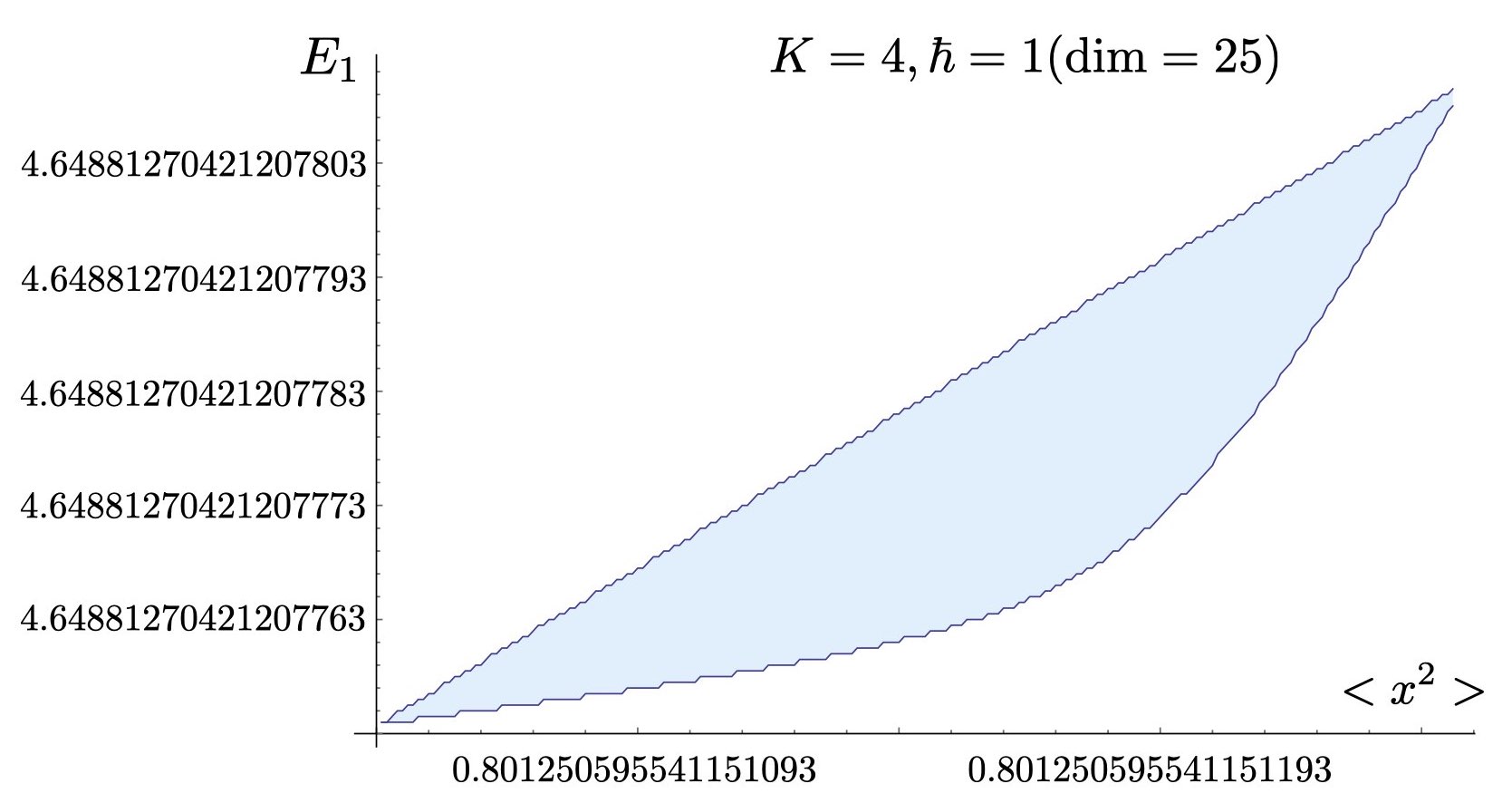}		
	\end{subfigure}

\caption{Bootstrap for the quartic anharmonic oscillator. We consider the ground state and first excited state, and plot the points which satisfy the bootstrap positivity constrains. For the first excited state, the bootstrap method does not give a good constrain at level $K=2$, so we only plot the levels $K=3,4$. }
 \label{x4diagram}
\end{figure}

We use the two-indexed operators $f_{m,n}$ for an improved bootstrap.  The relevant bootstrap relations are 
\be \label{bootstrap5.3}
	\bra H \hat{p}^m \hat{x}^n \ket = \bra \hat{p}^m \hat{x}^n H \ket = E \bra \hat{p}^m \hat{x}^n \ket.
\ee
One can move all the momentum operators $\hat{p}$ to the left to obtain the recursion relations for $f_{m,n}$'s. The useful formula is 
\be \label{recursionxp}
	\hat{x}^m \hat{p}^n = \hat{p} ( \hat{x}^m\hat{p}^{n-1} ) + i m \hbar  (\hat{x}^{m-1}\hat{p}^{n-1}) ,
\ee
which can be used inductively to write  $\hat{x}^m \hat{p}^n $ in terms of the operators $ \hat{p}^k \hat{x}^l $  with $k\leq n, l\leq m$. The relations are actually more complicated than those in the previous sections with exponential functions. Although we are not aware of a simple closed formula, it is a straightforward algorithm to obtain the expression of $\hat{x}^m \hat{p}^n $ for any small finite numbers $m,n$. 

We notice	that the Hamiltonian has the symmetry
\be	
	\hat{x} \rightarrow -\hat{x},~~ \hat{p} \rightarrow -\hat{p}. 
\ee
So if $m+n$ is odd, $f_{m,n}=0$. With the recursion relations from (\ref{bootstrap5.3}), we can solve for all $f_{m,n}$'s for even $m+n$ with the same initial conditions $E$ and $f_{0,2}$. One approach is to first solve for all the one-dimensional $f_{0,n}$'s and then solve for $f_{m,n}$'s inductively in $m$.  
	
Similarly, the Hermitian bootstrap matrix which satisfies positivity constrain is now 
\be
	M_{(m,n),(m',n')} =  \bra \hat{p}^m \hat{x}^{n+n'} \hat{p}^{m'} \ket. 
\ee
We can again move all momentum operators to the left and write the matrix elements as linear combinations of $f_{k,l}$'s. In Fig.  \ref{x4diagram} and Table \ref{x4eigenvalue}, we give some results of the bootstrap calculations about the ground state and the first excited state. The numerical accuracy of our two-dimensional approach is much better than that of the previous literature.

\begin{table}
\begin{center}

	\begin{tabular} {|c|c|c|} \hline  Bootstrap method  & $E_0 (\hbar=1)$ &   $E_1 (\hbar=1) $ \\
		  \hline  $K=2$          &  \underline{1.392}9548495694574126  &  \\
		  \hline  $K=3$          &  \underline{1.3923516415302}313493  & \underline{4.64881270}62682650584 \\
		  \hline  $K=4$          &  \underline{1.39235164153029185}49  & \underline{4.648812704212077}7642 \\
		  \hline \hline  Truncation method  &1.3923516415302918556                  &4.6488127042120775363\\
		  \hline
	\end{tabular}
	
	 \vskip 18pt
	 
	 \begin{tabular} {|c|c|c|} 
		 \hline  Bootstrap method & $\bra x^{2}\ket (\hbar=1)$ & $\bra x^{2} \ket (\hbar=1) $ \\
		 \hline   $K=2$           & \underline{0.305}72614942767230349  &\\
		  \hline  $K=3$           & \underline{0.30581365071}610749146  & \underline{0.801250595}91067376831 \\
		  \hline  $K=4$           & \underline{0.30581365071758713}570  & \underline{0.801250595541151}15879\\
		  \hline \hline  Truncation method &0.30581365071758713693                   &0.80125059554115104400\\
		  \hline
	 \end{tabular}
	
	\caption{The estimated values of the energy $E_0, E_1$ and $f_{0,2}$ for the quartic anharmonic oscillator. We average all the points in Fig. \ref{x4diagram} to get the estimated values, and compare them with the truncation method. The digits in the truncation method are stable with a truncation size of $300 \times 300$.  We underlie the digits in the bootstrap method which agree with the truncation method. }

\label{x4eigenvalue}
\end{center}

\end{table}

\section{Conclusion}  \label{conclusion} 

In our studies we mostly consider some fixed models without free parameters except for the Planck constant $\hbar$. It would be interesting to consider various deformations the models which can not be absorbed into the Planck constant, as well as the generalizations to more Calabi-Yau geometries. 

For the Calabi-Yau models in sections \ref{P1P1}, \ref{P2},  due to the constrains from the asymptotic behavior of the wave functions, there are only a finite number of available operators for bootstrap, where the bootstrap level is bounded by  
$\frac{\pi}{\hbar}$.  For the non-relativistic model in sections \ref{SU2} and \ref{secquartic}, we can in principle increase the bootstrap level without limit. In all cases, it appears that for the same computational level, the bootstrap method achieves much better accuracy when $\hbar$ is small, similarly as in the conventional truncation method.  It would be desirable to improve the situation so that the method can apply well to the case of large $\hbar$.

Comparing to the previous literatures in e.g. \cite{Han:2020bkb, Berenstein:2021dyf, Berenstein:2021loy, Bhattacharya:2021btd, Aikawa:2021eai, Aikawa:2021qbl}, we use a two-indexed operators with both $\hat{x}$ and $\hat{p}$. This turns out to improve greatly the efficiency of the bootstrap procedure. We can achieve quite high numerical precisions with only very low level $K\leq 4$, since the size of bootstrap matrix grows more quickly as $K^2$, comparing to the linear growth in $K$ in the previous  literatures. The exponential functions in our models also allow for both positive and negative indices in the cases with symmetry, e.g. the $\mathbb{P}^1\times \mathbb{P}^1$ model, further enlarging the bootstrap matrix.  We note that even with the same size of the bootstrap matrix, our two-dimensional prescription is probably still better since it can explore different corners of the region of the positivity constrains. 

In our scanned parameter space, near the physical exact values, the minimal eigenvalues of the bootstrap matrices are often very close to zero. In practice we do not exclude the points where they appear to be negative but the absolute values are reasonably small compared to the preset numerical accuracy, so that it is still possible that they can be actually positive. It would be better to have a more precise prescription for imposing the positivity constrain. 

It would be interesting to have a better understanding of the general pattern of the shapes of the allowed bootstrap regions, which often resemble narrow strips in our plotted figures. This would be helpful for a more efficient scan of the parameter space.  It is also interesting to consider the mathematical question whether these regions can in principle become arbitrarily infinitesimally small around the exact points, as we increase the size of bootstrap matrix asymptotically to infinity, as it is possible e.g. in the cases in sections \ref{SU2} and \ref{secquartic}. This is true in the case of the harmonic oscillator, that the bootstrap constrain gives the exact energy eigenvalues at a sufficiently large  level \cite{Aikawa:2021qbl}.

Overall, given the same amount of computational power, the numerical precisions of the bootstrap method in computing the energy eigenvalues in our models are still not better than those of the conventional truncation method. But of course, as a promising new development, the bootstrap method deserves to be further explored and improved to uncover its full potential.

\vspace{0.2in} {\leftline {\bf Acknowledgments}}
\nopagebreak

We thank Jun-hao Li, Gao-fu Ren for helpful discussions. This work was supported in parts by the National Natural Science Foundation of China (Grants  No.11947301 and No.12047502).

\appendix

\addcontentsline{toc}{section}{References}


\providecommand{\href}[2]{#2}\begingroup\raggedright\endgroup

\end{document}